\documentclass[manuscript]{aastex63}

\hypersetup{linkcolor=blue, urlcolor=blue}
\usepackage{graphicx}
\usepackage{amsmath}

\newcommand\solarmass{$M_\odot$}
\newcommand\vkm{km\,s$^{-1}$}
\newcommand\kms{km\,s$^{-1}$}

\submitjournal{AAS Journal}

\shorttitle{The Cassiopeia Filament}
\shortauthors{Chen et al.}

\begin{document}

\title{The Cassiopeia Filament: A Blown Spur of the Local Arm}

\correspondingauthor{Xuepeng Chen}
\email{xpchen@pmo.ac.cn}

\author{Xuepeng~Chen}
\author{Li~Sun}
\author{Jiancheng~Feng}
\author{Shiyu~Zhang}
\author{Weihua~Guo}
\author{Xiaoyun~Xu}
\affiliation{Purple Mountain Observatory \& Key Laboratory of Radio Astronomy, Chinese Academy of Sciences, 10 Yuanhua Road, 210023 Nanjing, China}
\affiliation{School of Astronomy and Space Science, University of Science and Technology of China, Hefei, Anhui 230026, China}

\author{Yang~Su}
\author{Yan~Sun}
\author{Shaobo~Zhang}
\author{Xin~Zhou}
\author{Zhiwei~Chen}
\author{Qing-Zeng~Yan}
\author{Miaomiao~Zhang}

\affiliation{Purple Mountain Observatory \& Key Laboratory of Radio Astronomy, Chinese Academy of Sciences, 10 Yuanhua Road, 210023 Nanjing, China}

\author{Min~Fang}
\author{Ji~Yang}
\affiliation{Purple Mountain Observatory \& Key Laboratory of Radio Astronomy, Chinese Academy of Sciences, 10 Yuanhua Road, 210023 Nanjing, China}
\affiliation{School of Astronomy and Space Science, University of Science and Technology of China, Hefei, Anhui 230026, China}

\begin{abstract}

We present wide-field and high-sensitivity CO\,(1--0) molecular line observations toward the Cassiopeia region, using the 13.7\,m millimeter telescope 
of the Purple Mountain Observatory (PMO). The CO observations reveal a large-scale highly filamentary molecular cloud within the Galactic region of 
132\fdg0\,$\geq$\,$l$\,$\geq$\,122\fdg0 and $-$1\fdg0\,$\leq$\,$b$\,$\leq$\,3\fdg0 and the velocity range from approximately +1 to $+$4\,\kms. The 
measured length of the large-scale filament, referred to as the Cassiopeia Filament, is $\sim$\,390\,pc. The observed properties of the Cassiopeia 
Filament, such as length, column density, and velocity gradient, are consistent with those synthetic large-scale filaments in the inter-arm regions. Based 
on its observed properties and location on the Galactic plane, we suggest that the Cassiopeia Filament is a spur of the Local arm, which is formed due to 
the galactic shear. The western end of the Cassiopeia Filament shows a giant arc-like molecular gas shell, which is extending in the velocity range from 
roughly $-$1 to $+$7\,\kms. Finger-like structures, with systematic velocity gradients, are detected in the shell. The CO kinematics suggest that the large 
shell is expanding at a velocity of $\sim$\,6.5\,\kms. Both the shell and finger-like structures outline a giant bubble with a radius of $\sim$\,16\,pc, which 
is likely produced by stellar wind from the progenitor star of a supernova remnant. The observed spectral linewidths suggest that the whole Cassiopeia 
Filament was quiescent initially until its west part was blown by stellar wind and became supersonically turbulent. 

\end{abstract}


\section{INTRODUCTION}

Multi-wavelength surveys found that filaments are ubiquitous in the molecular clouds (see, e.g., Schneider \& Elmegreen 1979; Molinari et al. 2010;
Li et al. 2016; Mattern et al. 2018; Yuan et al. 2021), which connect the collapse of the molecular clouds to the final fragmentation to form dense cores 
and thus play a key role in the process of star formation (see, e.g., Myers 2009; Andr\'e et al. 2014; Motte et al. 2018; Hacar et al. 2022). Furthermore, 
large-scale filaments, with lengths in the order of 100\,pc (see, e.g, Jackson et al. 2010; Ragan et al. 2014; Wang et al. 2015; Zucker et al. 2015; 
Abreu-Vicente et al. 2016; Colombo et al. 2021), are regarded as the ``feathers", ``spurs", or ``skeletons" of spiral arms, which are important for 
understanding the Galactic structure (see, e.g., Goodman et al. 2014; Smith et al. 2014, 2020; Zucker et al. 2019).

Bubbles and/or shells are also ubiquitous in the multi-wavelength surveys toward the Galactic plane (see, e.g., Churchwell et al. 2006, 2007; Simpson 
et al. 2012). Most of these bubbles, with radii ranging from a few to tens of pc, are created by the feedback from massive stars, such as UV radiation (e.g., 
Churchwell 2002; Deharveng et al. 2010) and/or stellar winds (e.g., Weaver et al. 1977; Freyer et al. 2003; Arthur 2007). Even larger structures with 
radii $\geq$\,100\,pc, the so-called superbubbles or supershells, are suggested to be produced by OB associations and/or supernovae in the Milky Way 
(e.g., McClure-Griffiths et al. 2002; Suad et al. 2014; Wright 2020; Zucker et al. 2022). 

As both filaments and bubbles are ubiquitous in the Milky Way, the interaction between these two structures can be expected then. Indeed, previous 
observations found a few cases. Depending on the formation sequence and environment, the interaction may be divided into three different 
types:
(a) Filaments are affected by bubbles produced by externally pre-existing stars. For example, a large-scale filamentary wisp in the Galactic region of 
52\fdg5\,$>$\,$l$\,$>$\,49\fdg5 is disturbed by a giant bubble, which is likely driven by the expansion of a supernova (see Li et al. 2013).
(b) Filaments are interacted with bubbles created by stars formed therein, such as the N131 bubble (Zhang et al. 2016) and the ``Nessie'' filament 
(Jackson et al. 2021).
(c) Filaments are formed through converging flows caused by (super-)\,bubbles (e.g., Inutsuka et al. 2015), and then continuously affected by parent
bubbles. The filament B211/B213 in Taurus could be a promising case for this kind of interaction (see Shimajiri et al. 2019).

Wide-field observations of molecular lines provide physical, chemical and dynamic information of molecular gas, which is essential to study the interaction 
between large-scale filaments and bubbles. This kind of observations can help us to better understand the formation mechanism and kinematics of filamentary 
molecular clouds, as well as the feedback from stars on the interstellar medium (ISM).
The Milky Way Imaging Scroll Painting (MWISP) project\footnote{http://www.radioast.nsdc.cn/mwisp.php} is an unbiased and sensitive CO\,(1--0) multi-line 
survey toward the northern Galactic plane, using the 13.7\,m millimeter telescope of the Purple Mountain Observatory (PMO; Su et al. 2019; Sun et al. 2021),
which provides high-quality data for studying the distribution and property of molecular gas in the Milky Way.
As part of the MWISP survey, we present in this work wide-field CO\,(1--0) observations toward the Cassiopeia region. Based on the CO observations, we 
report the discovery of a large-scale spur-like molecular filament in the Local arm, which is likely interacting with a giant bubble in the region.
In Section 2 we describe the observations and data reduction. Observational results are presented in Section 3 and discussed in Section 4. 
The main conclusions of this study are summarized in Section 5.

\section{OBSERVATIONS AND DATA REDUCTION}

The CO\,(1--0) observations toward the Cassiopeia region were performed from November 2011 to September 2020 with the PMO 13.7\,m 
telescope at Delingha in China. The nine-beam Superconducting Spectroscopic Array Receiver (SSAR; see Shan et al. 2012) was used at 
the front end in the sideband separation mode in the observations. Three CO\,(1--0) lines were simultaneously observed, $^{12}$CO at 
the upper sideband (USB) and two other lines, $^{13}$CO and C$^{18}$O, at the lower sideband (LSB). A Fast Fourier Transform (FFT) 
spectrometer with a total bandwidth of 1\,GHz and 16,384 channels was used as the back end. The corresponding velocity resolutions were 
$\sim$\,0.16\,km\,s$^{-1}$ for the $^{12}$CO line and $\sim$\,0.17\,km\,s$^{-1}$ for both the $^{13}$CO and C$^{18}$O lines. 
The observed region was divided into individual 30$'$\,$\times$\,30$'$ cells in the observations. Each cell was mapped with the On-The-Fly 
(OTF) mode. In order to reduce scanning effects, each cell was mapped at least twice, along the Galactic longitude and latitude, respectively. 
The half-power beam width (HPBW) was $\sim$\,52$''$ for the $^{12}$CO line and $\sim$\,55$''$ for the $^{13}$CO and C$^{18}$O lines, 
while the pointing accuracy was $\sim$\,5$''$.

The standard chopper-wheel method was used to calibrate the antenna temperature ($T_{\rm A}$). The relationship $T_{\rm mb}$ = $T_{\rm A}$/$B_{\rm eff}$ 
was used to convert the antenna temperature $T_{\rm A}$ to the main-beam temperature ($T_{\rm mb}$), where the main-beam efficiencies ($B_{\rm eff}$) 
during the observations were $\sim$\,44\% for the USB and $\sim$\,48\% for the LSB, respectively. The calibration errors were estimated to be within 10\%. 
During the observations, typical system temperatures were around 210\,K for the USB and around 130\,K for the LSB, and the variations among 
different beams were less than 15\%. 

After removing bad channels and abnormal spectra, and correcting the first order (linear) baseline fitting, the data were re-gridded into standard 
FITS files with a pixel size of 30$\arcsec$\,$\times$\,30$\arcsec$ (approximately half of the beam size).  The average rms noises of all final spectra 
were about 0.5\,K for $^{12}$CO (at a velocity resolution of $\sim$\,0.16\,\kms) and about 0.3\,K for $^{13}$CO and C$^{18}$O (at a velocity resolution 
of $\sim$\,0.17\,\kms). Velocities were all given with respect to the local standard of rest (LSR) in this work. Finally, we mosaicked the data cubes 
toward the Cassiopeia region to analyze the distribution and properties of molecular gas. All data were reduced using the GILDAS package (see 
https://www.iram.fr/IRAMFR/GILDAS/).

\section{RESULTS}

The MWISP CO data toward the Galactic Cassiopeia region were partially presented by Sun et al. (2021; 141\fdg54\,$\geq$\,$l$\,$\geq$\,104\fdg75
and $-$3\fdg028\,$\leq$\,$b$\,$\leq$\,5\fdg007) and Yan et al. (2021; 150\fdg25\,$\geq$\,$l$\,$\geq$\,104\fdg75 and $|b|$\,$\leq$\,5\fdg25), respectively. 
The work by Sun et al. (2021) provided global properties of the MWISP CO data and examined CO completeness through a comparison between three 
independent CO surveys, while Yan et al. (2021) measured distances toward 76 medium-sized molecular clouds in the second Galactic quadrant.
In this work, we investigate the Galactic region of 135\fdg0\,$\geq$\,$l$\,$\geq$\,120\fdg0 and $-$3\fdg5\,$\leq$\,$b$\,$\leq$\,4\fdg5 and the velocity 
range between $-$5 and +10\,\kms, and present insights into the structure, physical properties, and kinematics of the molecular gas therein.

\subsection{A Giant Molecular Filament Associated with A Shell} 

Figure~1 shows the velocity channel maps of the $^{12}$CO\,(1--0) emission toward the observed region. As seen in the channel maps, a large-scale
highly filamentary molecular cloud is detected in the velocity range between about +1 and +4\,\kms. This filamentary cloud extends roughly 10 degrees 
along the longitude direction (from $\sim$\,122$^\circ$ to $\sim$\,132$^\circ$), with an inclination angle of $\sim$\,20$^\circ$ to the Galactic plane. 
Figure~2 shows the velocity channel maps of the $^{13}$CO\,(1--0) emission, which traces dense molecular gas along the filamentary cloud. 

Figure~3 shows the CO velocity-integrated intensity images. We adopt the Discrete Persistent Structures Extractor (DisPerSE) algorithm (Sousbie 2011), 
a method for identifying filamentary structure in the ISM, to outline filaments in the $^{12}$CO image. 
After testing several different combinations between Persistence and Robustness threshold 
(see DisPerSE website\footnote{http://www2.iap.fr/users/sousbie/web/html/index55a0.html?category/Quick-start} for more details), we set the value of 
10\,K\,\kms\ ($\sim$\,17\,$\sigma$) for persistence and 1\, K\,\kms\ ($\sim$\,2\,$\sigma$) for robustness. This setting would not only guarantee a clear 
skeleton construction with a high contrast with the surroundings but also retain the information of persistence in morphology as far as possible.
The blue line in Figure~3 shows the main filament identified by the DisPerSE algorithm, which is consistent with the large-scale filament found by visual 
inspection in the $^{12}$CO channel maps. The aspect ratio of the main filament is measured to be $\sim$\,28 in the intensity image.
Several sub-filaments are also identified by DisPerSE, which are shown by the orange lines in Figure 3. These sub-filaments are roughly perpendicular 
to the main filament, which together may be morphologically interpreted in terms of a hub-filament system (see, e.g., Myers 2009).
Hereafter we refer to the large-scale main filament as the Cassiopeia Filament and focus on it in this work.

Interestingly, as seen in Figure~3, the western part of the Cassiopeia Filament presents a clear arc-like shell\footnote{We note that a bubble-like structure 
is seen in the east of the main filament (see Figures~1\,\&\,3). After checking the CO spectra, we find that molecular gas therein is quiescent ($^{12}$CO linewidth 
less than 1\,\kms). No expansion signature is seen in the CO position-velocity diagrams either. Therefore, we do not further discuss this structure in the work.}, 
which opens to the north and covers an extent with coordinates of 125\fdg5\,$\geq$\,$l$\,$\geq$\,121\fdg5 and $-$1\fdg3\,$\leq$\,$b$\,$\leq$\,0\fdg5. Figure~4 
shows the enlarged $^{12}$CO velocity channel maps for this western shell. The shell appears at the velocity of about $-$1\,\kms\ and fades away at the 
velocity of $\sim$\,+7\,\kms, with its peak emission at the velocity of $\sim$\,3-4\,\kms.
Figure~5 shows the $^{12}$CO intensity image integrated over the velocity range from $-$1 to +7\,\kms. Finger-like structures, all pointing toward south, 
are clearly seen in this shell, which are named after numbers according to their positions from west to east (see Fig.\,5). The arc-like shell, as well as the 
finger-like structures on the shell, are also clearly detected in the $^{13}$CO images (see Figure~16 below).
We also note that no C$^{18}$O\,(1--0) emission is detected from the main filament and its associated shell, except the tips of Finger~1 ($\sim$\,7\,$\sigma$) 
and Finger~4 ($\sim$\,5\,$\sigma$) (see Figure~6).

\subsection{The Length and Mass of the Cassiopeia Filament} 

The distance, which is a fundamental parameter for deriving physical properties, is unclear yet for the Cassiopeia Filament. We note that the LSR 
velocity ranges of spiral arms in the observed direction are about [$-$60, $-$30]\,\kms\ for the Perseus arm and [$-$25, +10]\,\kms\ for the Local arm, 
respectively (see Dame et al. 2001; Sun et al. 2021). The CO velocity range of the Cassiopeia Filament and its associated shell is roughly from $-$1 
to +7\,\kms, which is covered by the velocity range of the Local arm.

Based on the MWISP CO and $Gaia$ DR2 data (Gaia et al. 2018), we measure the distance toward the Cassiopeia Filament, using the same method 
as described in Yan et al. (2021). This method uses the principle that molecular clouds usually impose higher optical extinction than other phases of 
the ISM. Aided by Bayesian analyses, we derive distances by identifying the breakpoint in the stellar extinction toward molecular cloud (on-cloud region) 
and using the extinction of $Gaia$ stars around molecular cloud (off-cloud regions) to confirm the breakpoint. The systematic error in the measurement 
is approximately 5\%.

The Cassiopeia Filament extends roughly 10$^\circ$ along the longitude direction. Therefore, the distance along the filament may be changed largely,
suggested by the pattern of the Local arm in this direction (see, e.g., Reid et al. 2019; Xu et al. 2021). Indeed, the measured distance to the east filament 
end is roughly 188$^{+5}_{-8}$\,pc (see Figure~7a), while the distance to the west shell is about 553$^{+8}_{-9}$\,pc (Figure~7b). 
Taking 10$^\circ$ as the viewing angle (on the Galactic plane) and 20$^\circ$ as the inclination angle (to the Galactic plane), the calculated length of 
the Cassiopeia Filament is roughly 390\,pc.

Two methods have been used to derive the H$_{2}$ gas column density. In the first method, on the assumption of local thermodynamic 
equilibrium (LTE) and optically thick $^{12}$CO line ($\tau_{^{12} \rm CO}$ $\gg$ 1), we can derive the excitation temperature $T_{\rm ex}$ 
from the peak radiation temperature of the $^{12}$CO line by following formula:
\begin{equation}\label{f.Tex}
T_{\rm mb} = [J(T_{\rm ex}) - J(T_{\rm bg})][1 - e^{(-\tau)}],
\end{equation}
where $T_{\rm mb}$ is the main-beam temperature, $T_{\rm bg}$ is the background temperature with the value of 2.7\,K; 
$J_{\nu}$~=~$T_0$/[$e^{(h \nu / k_{\rm B}T)}$ $-$ 1], here $T_0$ is the intrinsic temperature of $^{12}$CO and $T_0$ = $h \nu / k_{\rm B}T$,
$k_{\rm B}$ is the Boltzmann constant, $h$ is the Planck constant.

Assuming an equal $T_{\rm ex}$ of the isotopic pair, the $^{13}$CO column density $N_{\rm ^{13}CO}$ can be estimated over the 
velocity ($V$) by
\begin{equation}\label{f.13CO}
N_{\rm ^{13}CO} = 2.42 \times 10^{14} \times \frac {(1 + 0.88/T_{\rm ex}) \times \int T_{\rm mb, ^{13}CO} \, dV} {1 - e^{-T_0 (^{13}{\rm CO})/T_{\rm ex}}}.
\end{equation}
The H$_2$ column density could then be derived by multiplying the $^{13}$CO column density by the ratio of 
$N_{\rm H_2}$/$N_{\rm ^{13}CO}$\,$\sim$\,7\,$\times$\,10$^5$ (Frerking et al. 1982).
In the second method, the H$_2$ column density can be estimated by integrating the main-beam temperature of $^{12}$CO over the velocity, using 
a mean CO-to-H$_2$ conversion factor $X$ = 1.8\,$\times$\,10$^{20}$\,cm$^{-2}$\,K$^{-1}$\,km$^{-1}$\,s (Dame et al. 2001):
\begin{equation}\label{f.H2}
N_{\rm H_2} = X  \int T_{\rm mb, ^{12}CO} \, dV.
\end{equation}
Figure~8 shows the distributions of excitation temperature and H$_2$ column densities derived from above methods for the Cassiopeia Filament.

The gas mass can then be derived from the H$_{2}$ column density $N_{\rm H_2}$ by 
\begin{equation}\label{f.mass}
 M = \mu m_{\rm H} D^{2}  \int N_{\rm H_2}{\rm d}\Omega\;,
\end{equation}
where $\mu$ is the mean molecular weight per hydrogen molecule which is assumed to be 2.8, $m_{\rm H}$ is the mass of the atomic hydrogen, 
$D$ is the distance to molecular cloud, {\rm d}$\Omega$ is the solid angle element.
For the arc-like shell, we adopt the distance of 553\,pc, and the derived mass is roughly $\sim$\,1200\,$M_\odot$ (LTE method) and $\sim$\,6100\,$M_\odot$ 
($X$-factor method), respectively. For the rest filamentary cloud, a mean distance of 400\,pc is adopted, and the estimated mass is roughly $\sim$\,860\,$M_\odot$ 
(LTE method) and $\sim$\,8300\,$M_\odot$ ($X$-factor method), respectively. 
The difference between the masses derived by the two methods is mainly caused by the small filling factors of the $^{13}$ CO emission, because 
there is much less $^{13}$CO emission than $^{12}$CO emission in the cloud (see Figures~3\,\&\,8).

\subsection{The Gas Kinematics of the Cassiopeia Filament} 

Figure~9 shows the $^{12}$CO intensity-weighted velocity field (1st moment image) of the observed region, where the distribution of the mean 
velocities ($V_{\rm LSR}$) of the molecular gas can been seen. As seen in Fig.\,9, the Cassiopeia Filament shows a velocity gradient increasing 
from the east to the west. This velocity gradient is also seen in the CO position-velocity (PV) diagram along the filament shown in Figure~10. The
whole Cassiopeia Filament is velocity-coherent and the measured velocity gradient is roughly 0.4\,\kms\,deg$^{-1}$ or $\sim$\,0.013\,\kms\,pc$^{-1}$ 
(adopting a filament length of 390\,pc). 

Figure~9 also shows the velocity field for the shell. As seen in the image, the shell also shows clear systematic velocity gradients but increasing from 
north to south. The gradients are much clearer in the finger-like structures, in particular, Fingers~ 1, 2, 4 and 7 (see Fig.\,9). The measured typical 
velocity gradient is roughly 1.0\,$\pm$\,0.2\,\kms\,pc$^{-1}$ at the distance of 553\,pc for the shell. The PV diagrams of these finger-like structures show 
curve-shaped morphologies at their tips. This feature is most obvious in Fingers~2 and 4, which could be well fitted by the ellipses with velocity radii 
of $\sim$\,6.5\,$\pm$\,0.5\,\kms\ (see Fig.\,11). The results imply that the shell is expanding. 

Figure~12 shows the distribution of the linewidths (2nd moment image) in the CO observations. For a molecular cloud, the thermal linewidth is 
calculated as $\Delta v_{\rm thermal}$ = $\sqrt{8{\rm ln}2 \frac{k_{\rm B}T_{\rm kin}}{\mu_{\rm obs}m_{\rm H}}}$, where $k_{\rm B}$ is the Boltzmann 
constant and $\mu_{\rm obs}$ is the molecular weight of the observed species (2.33 for H$_2$). Adopting the kinetic temperature $T_{\rm kin}$ of 10\,K 
(similar to excitation temperature; see Fig.\,8a), the $\Delta v_{\rm thermal}$ is estimated to be $\sim$\,0.45\,\kms\ for the Cassiopeia Filament. The 
non-thermal linewidth $\Delta v_{\rm NT}$ is calculated as $\sqrt{\Delta v_{\rm obs}^2 - \Delta v_{\rm thermal}^2 - \Delta v_{\rm res}^2}$, where 
$\Delta v_{\rm obs}$ is measured FWHM linewidth of the observed spectra and $\Delta v_{\rm res}$ is the velocity resolution in the observations 
(0.16\,\vkm\ for $^{12}$CO).
For the east filament, the observed $^{12}$CO linewidths range from $\sim$\,0.8 to $\sim$\,1.1\,\kms, with a median value of $\sim$\,0.9\,\kms\ (see also
PV diagram shown in Fig.\,10). The non-thermal linewidths are then calculated to be about 0.6--1.0\,\kms, with a median value of $\sim$\,0.8\,\kms. For 
the west shell, the observed $^{12}$CO linewidths are roughly 2-3\,\kms, and the calculated non-thermal linewidths are in the same range. Generally, 
the ratio of  $\Delta v_{\rm NT}$/$\Delta v_{\rm thermal}$ is defined as the observed Mach number $M$, which is used to distinguish between the subsonic 
($M$\,$<$\,1), transonic (1\,$\leq$\,$M$\,$\leq$\,2), and supersonic ($M$\,$>$\,2) hydrodynamical regimes in isothermal, non-magnetic fluids. Therefore, 
in the $^{12}$CO observations toward the Cassiopeia Filament, its east filament is transonic, while the west shell is supersonic. We also note that the 
observed $^{13}$CO linewidths in the east filament are roughly 0.5--0.6\,\kms, and the calculated non-thermal linewidths are about 0.2--0.4\,\kms. 
Therefore, the east filament is subsonic in the $^{13}$CO observations (see discussion in $\S$\,4.3.2). 

The line mass $M_{\rm line}$ of the Cassiopeia Filament is $\sim$\,37\,\solarmass\,pc$^{-1}$, estimated from the gas mass ($\sim$\,14400\,\solarmass; 
$X$-factor method) and length ($\sim$\,390\,pc). The critical line mass of an unmagnetized isothermal filament can be described in the following form 
(Ostriker 1964; Inutsuka \& Miyama 1997):  $M_{\rm crit}$ = 2$c_{\rm s}^2$/$G$ $\approx$ 16.4\,$\times$\,($\frac{T}{10\,{\rm K}}$)\,\solarmass\,pc$^{-1}$, 
where $c_{\rm s}$ = $\sqrt{ \frac{k_{\rm B}T_{\rm kin}}{\mu_{\rm obs}m_{\rm H}}}$  is local thermal sound speed ($\sim$\,0.19\,\kms\ at 10\,K) and $G$ is 
the gravitational constant. For the Cassiopeia Filament, the $M_{\rm crit}$ is estimated to be $\sim$\,16.4\,\solarmass\,pc$^{-1}$. 
As line mass $M_{\rm line}$ is larger than critical equilibrium mass $M_{\rm crit}$,  the Cassiopeia Filament is in thermally supercritical state, which will be 
gravitationally unstable to collapse radially and subject to fragmentation along the length (see Inutsuka \& Miyama 1997). 

\section{DISCUSSION}

\subsection{The Origin of the Cassiopeia Filament: A Spur of the Local Arm}

Large-scale molecular filaments, with lengths in order of 100\,pc, play an important role in both star formation and Galactic structure. A number of 
large-scale molecular filaments were found in the past decade (see, e.g., Jackson et al. 2010; Li et al. 2013; Ragan et al. 2014; Zucker et al. 2015; 
Wang et al. 2015, 2016; Abreu-Vicente et al. 2016; Li et al. 2016; Du et al. 2017; Guo et al. 2021; Colombo et al. 2021). Depending on searching 
tracers and identifying methods, these filaments have various properties and can be generally divided into `giant molecular filaments' (GMFs; e.g., 
Ragan et al. 2014; Abreu-Vicente et al. 2016) and `bone-like' filaments (e.g., Wang et al. 2015; Zucker et al. 2015) two groups\footnote{In Zucker 
et al. (2018), four catalogs, i.e., giant molecular filaments (from Ragan et al. 2014 and Abreu-Vicente et al. 2016), Milky Way bones (from Zucker 
et al. 2015), large-scale Herschel filaments (from Wang et a. 2015), and MST bones (from Wang et al. 2016), are compared. The `bone-like' filaments, 
including both Milky Way bones and large-scale Herschel filaments, have similar physical properties (e.g., column density and aspect ratio), which 
are clearly different from those of the giant molecular filaments. As discussed in Zucker et al. 2018, the MST bones could be elongated dense core 
complexes tracing out networks of dense compact sources embedded in giant molecular clouds.} (see properties summarized by Zucker et al. 2018).

The measured length of the Cassiopeia Filament is roughly 390\,pc, which is comparable to the 500\,pc wisp found by Li et al. (2013) and the 430\,pc 
``Nessie Optimistic" found by Goodman et al. (2014). It then belongs to the longest molecular filaments in the Galaxy. Its relative low H$_2$ column 
density ($\sim$\,2\,$\times$\,10$^{21}$ cm$^{-2}$) is comparable to those of GMFs (with a median value of 4.8\,$\times$\,10$^{21}$ cm$^{-2}$), but 
much lower than those of `bone-like' filaments ($\sim$\,7.8-10.0\,$\times$\,10$^{21}$ cm$^{-2}$). We note that the aspect ratio of the Cassiopeia Filament 
($\sim$\,28) is larger than those of GMFs (with a median value of $\sim$\,8) and similar to `bone-like' filaments (with a median value of $\sim$\,34; see 
Zucker et al. 2018). 

The line mass of the Cassiopeia Filament is $\sim$\,37\,\solarmass\,pc$^{-1}$ (see $\S$\,3.3). This value is much smaller than those of the GMFs 
(with a median value of $\sim$\,1500\,\solarmass\,pc$^{-1}$) and `bone-like' filaments ($\sim$\,500\,\solarmass\,pc$^{-1}$). Colombo et al. (2021) 
recently identified a sample of outer Galaxy large-scale filaments (OGLSF), whose physical properties  (such as low gas masses and low line masses) 
can be distinguished from  those of the GMFs and `bone-like' filaments found in the inner Galaxy. As suggested by Colombo et al. (2021), these OGLSF 
are located within inter-arm regions. Interestingly, the observed line mass of the Cassiopeia Filament is more similar to the line masses of the OGLSF 
($\sim$\,10--600\,\solarmass\,pc$^{-1}$, with a median value of $\sim$\,50\solarmass\,pc$^{-1}$; see Colombo et al. 2021).

The formation of large-scale filaments is still in study. Several formation mechanisms are suggested, such as galactic dynamics, including galactic potential
in spiral arms and differential shear in the inter-arm regions (see, e.g., Smith et al. 2014, 2020; Duarte-Cabral \& Dobbs 2016, 2017; Zucker et al. 2019), and 
localized dynamics, including self-gravity (e.g., Burkert \& Hartmann 2004; Hartmann \& Burkert 2007; Gomez \& Vazquez-Semadeni 2014), stellar feedback 
(e.g., Inutsuka et al. 2015), or supersonic turbulence (e.g., Ballesteros-Paredes et al. 1999; Hennebelle 2013).

Based on the observed velocity range and estimated distance, we suggest that the Cassiopeia Filament is associated with the Local arm. For comparison, 
the FWHM thickness of molecular disk at the Galactocentric radius of about 8\,kpc is roughly 100--200\,pc (see Heyer \& Dame 2015), while the width of the 
Local arm is roughly $\sim$\,300\,pc (see Reid et al. 2019). Therefore, we consider that the Cassiopeia Filament, with a length of $\sim$\,390\,pc, is difficult 
to be formed by localized mechanisms within the Local arm. On the other hand, the observed properties of the Cassiopeia Filament, such as large length, 
low column density ($\sim$\,2\,$\times$\,10$^{21}$ cm$^{-2}$), and small velocity gradient ($\sim$\,0.013\,\kms\,pc$^{-1}$), are consistent with those 
synthetic large-scale molecular filaments in the inter-arm regions from the galactic-scale simulations (e.g., Smith et al. 2014; Duarte-Cabral \& Dobbs 2017; 
Zucker et al. 2019). Furthermore, its observed low line mass is comparable to those large-scale filaments found in the outer Galaxy inter-arm regions 
(see Colombo et al. 2021).
Figure~13 shows multi-scale CO longitude-velocity diagrams in the observed direction. The velocity range of the Cassiopeia Filament ([+1, +4]\,\kms) is 
at the edge of the velocity range of the Local arm, roughly 10\,\kms\ away from the central mean velocity of the arm (about --8\,\kms; see Figure~13). Figure~14 
shows the position of the Cassiopeia Filament with respect to the Galactic molecular gas disk. It is seen that the filament is vertically located at the boundary 
of the disk, which is different from those `bone-like' filaments found in the mid-plane of the inner Galaxy.
Figure~15 shows the projected location of the Cassiopeia Filament on the plane of the Milky Way (face-on view; see Reid et al. 2019). As seen in Figure~15, 
the filament is just like a spur of the Local arm. Based on the observed properties and location, we then suggest that the Cassiopeia Filament is formed due 
to the galactic shear.

\subsection{The Formation of the Shell: A Giant Wind-Blown Bubble}

The observed morphology and kinematics suggest that the arc-like shell at the west end of the Cassiopeia Filament is resulted from the expansion of 
a large bubble. Under this assumption, we try to search for the center of the bubble by measuring the convergence point of the finger-like structures 
(see Figure~16). The measured geometric center is $l$\,=\,124\fdg06 and $b$\,=\,1\fdg15, with an error radius of $\sim$\,0\fdg20. The estimated dimension 
of the bubble is about 3\fdg4\,$\times$\,2\fdg8, corresponding to $\sim$\,34\,pc\,$\times$\,28\,pc at the distance of 553\,pc.

It is well-known that OB stars are able to produce bubbles in the ISM through strong feedback, such as UV radiation, stellar winds, and supernova 
explosions  (see, e.g., reviews by Krumholz et al. 2014 and Dale 2015). To better understand the interstellar environment of the shell, we show in 
Figure~16 the MWISP $^{12}$CO contours, plotted on the Urumqi 6\,cm radio continuum image from Sun et al. (2007).
As seen in Figure~16, there are several known HII regions toward the bubble suggested by the shell, such as S183 (to the northeast of the shell), S185 
(to the south of the shell), and S186 (to the east of the shell). Nevertheless, radio recombination line observations suggested that the systemic velocity 
of S183 is about $-$63\,\kms\ (Landecker et al. 1992); The distance of S185 was estimated to be $\sim$\,200\,pc (e.g., Soam et al. 2017); For S186, the 
systemic velocity of the CO gas was found to be about $-$43\,\kms\ (Qin et al. 2008). Hence, we could exclude the association between these HII regions 
and the shell found in this work. We further checked archival H$\alpha$ images (not shown here), but did not find any ionized structure spatially coincident 
with the molecular gas shell. Therefore, we consider that the gas shell is not resulted from the expansion of an HII region.

As found by Chen et al. (2013), there is a linear relationship between the radius of a wind-blown bubble in a molecular cloud ($R_{\rm b}$) and the initial 
mass of the energy source star ($M_{\rm star}$): $R_{\rm b}$ (pc) $\approx$ 1.22$M_{\rm star}$/$M_\odot$ $-$ 9.16\,pc, assuming a constant inter-clump 
pressure (see Chen et al. 2013 for more details). For the giant bubble suggested by the shell (effective radius of $\sim$\,16\,pc), a massive star with a 
mass of $\sim$\,21\,$M_\odot$ (O7 or earlier types) is required to supply stellar wind. Previous observations showed a stellar cluster, Cassiopeia OB7, 
toward the center of the giant bubble (see Garmany \& Stencel 1992; Cazzolato \& Pineault 2003). Nevertheless, the distance of the Cassiopeia OB7 
cluster is about 2\,kpc (e.g., Cazzolato \& Pineault 2003), far away from the shell. Therefore, we consider no association between the Cassiopeia OB7 
cluster and the shell/bubble. We also note that no O-type stars were found yet in the bubble/shell region in the Local arm (see, e.g., Xu et al. 2021).
Based on the $Gaia$ data, we further search for new OB stars in the shell region (see Appendix~A). However, we do not find any candidates associated 
with the shell.

We investigated known supernova remnants (SNRs) in this Galactic region. Large shell-type SNRs G126.2+1.6 and G127.1+0.5 (see Zhou et al. 2014) 
were found to the east of the bubble (see Fig.\,16). Nevertheless, we did not find any known SNRs enclosed by the giant bubble. As seen in Figure~16, a faint 
and extended radio continuum source, with a dimension of $\sim$\,55$'$\,$\times$\,60$'$, is detected toward the center of the giant bubble in the Urumqi 6\,cm 
radio continuum survey (see Sun et al. 2007), which was named after G124.0+1.4. Sun et al. (2007) suggested that this extended radio source could be an 
HII region, as a B-type star was found at the center of the extended radio source\footnote{After checking the SIMBAD Astronomical Database 
(http://simbad.cds.unistra.fr/simbad/), we found two Be stars, 2MASS J01005124+6413271 and 2MASS J01013802+6413500, located in the center of the radio 
source G124.0+1.4. The parallaxes of the two stars, measured by the $Gaia$ observations, are 0.3086 and 0.3097, respectively.}. If this extended radio source 
is really associated with the B-type stars in the center, its distance should be more than 3\,kpc. On the other hand, we note that the spectra index of the radio source 
G124.0+1.4 is about $-$0.22 (see Sun et al. 2007). This index is comparable with those of SNRs measured in the Urumqi 6\,cm radio continuum survey (Sun et 
al. 2011; see also a review by Dubner \& Giacani 2015). This implies that G124.0+1.4 might be a candidate SNR.

We show multi-wavelength archived radio continuum images of G124.0+1.4 in Figure~17, in order to better understand its nature. After comparing the Green 
Bank 4850\,MHz (6\,cm) image with the Urumqi 6\,cm image (see Fig.\,17a), we find that the extended radio emission of G124.0+1.4 is actually dominated  
by three radio sources, in which source 1 is spatially confident with the bubble center referred from the CO shell and fingers. The three radio sources are also 
detected in the low-frequency radio observations. The measured fluxes of the three sources are listed in Table~1. The fitted spectra indexes indicate that the radio
emission of the three sources is non-thermal. The results further support that G124.0+1.4 is a candidate SNR. In this picture, the giant bubble could be produced by 
the stellar wind from the progenitor star of this SNR.

In a short summary, based on the CO morphology and kinematics, we suggest that the arc-like shell is originally a part of the Cassiopeia Filament, which is 
blown into current morphology by strong stellar wind. The wind is likely driven by the progenitor star of an SNR in the region. Further observations are needed 
to search for massive stars in the bubble region on the one hand and to study the nature of radio source G120.0+1.4 on the other hand, in order to better 
understand the driving source of this expanding shell.

\subsection{The Origin of Non-thermal Linewidths: from Sonic to Supersonic}

The observed spectral linewidths in the molecular clouds are generally larger than thermal linewidths (see, e.g., reviews by Dobbs et al. 2014; 
Heyer \& Dame 2015; Ballesteros-Paredes et al. 2020). Taking the Cassiopeia Filament as an example, the observed $^{12}$CO non-thermal
linewidths are about 0.6--1.0\,\kms\ (with a median value of $\sim$\,0.8\,\kms) in the east filament and about 2--3\,\kms\ in the west shell (see 
$\S$\,3.3). The observed $^{12}$CO non-thermal linewidths are larger than the thermal linewidth ($\sim$\,0.45\,\kms) in the Cassiopeia Filament.

The origin of non-thermal linewidths in the molecular clouds is still in debate. Several scenarios, such as global collapse, externally-driven turbulence, 
and internally-driven turbulence, were suggested (see, e.g., Dobbs et al. 2014). Numerous observations and simulations find that stellar feedback 
(including outflows, HII regions, winds, and supernovae), no matter external or internal to clouds, can drive turbulence and explain large linewidths 
(Dobbs et al. 2014; Krumholz et al. 2014). By a comparison between the east filament and west shell, we suggest that large non-thermal linewidths 
seen in the shell are resulted from stellar wind, which blew the west end of the Cassiopeia Filament and also drove supersonic turbulence in the 
shaped shell. 

It is of interest to note that the east filament is transonic in the $^{12}$CO observations and subsonic in the $^{13}$CO observations (see $\S$\,3.3).
This means that thermal motion plays an important role in the east filament\footnote{The angular length of the east filament is roughly 9$^\circ$ (see,
e.g., Figure~10), corresponding to a linear length of $\sim$\,65\,$\times$\,$\frac{d}{400}$\,pc (where a mean distance of 400\,pc is adopted in this 
work).}, while the non-thermal linewidths could be resulted from gravitational radial collapse or potential star formation (e.g., mass accretion) along 
the filament\footnote{Indeed, several prestellar and protostellar sources were detected along the filament (see Figure~12) in the Herschel Hi-Gal 
survey (Elia et al. 2021). With similar molecular gas velocities, these Hi-Gal sources are likely associated with the filament, indicating early phase 
star formation therein.}. Based on the observations, we suggest that the whole Cassiopeia Filament could be quiescent initially until its west part was 
blown by stellar wind. Compared with other large-scale molecular filaments (see, e.g., Li et al. 2013; Ragan et al. 2014; Wang et al. 2015; Zucker et al. 
2015; Abreu-Vicente et al. 2016), the Cassiopeia Filament stands out by its large length and small linewidths. 
This kind of quiescent and long molecular filaments is rare in the observations. Another well-known example is Musca, a nearby $\sim$\,6\,pc-long 
filament with low column density ($\sim$\,2\,$\times$\,10$^{21}$\,cm$^{-2}$; Cox et al. 2016) and transonic/subsonic gas motion in the $^{13}$CO/C$^{18}$O 
observations (Hacar et al. 2016). 

As found in Zhang et al. (2019), there is a scaling relation between linewidth ($\Delta v$) and size ($R$): $\Delta v$\,$\propto$\,$R^{0.42\,\pm\,0.1}$, in 
the statistic toward the filaments with lengths ranging from $\sim$\,10\,pc to $\sim$\,100\,pc (see also a recent review by Hacar et al. 2022). This relation 
is reminiscent of the linewidth-size relation observed in the molecular clouds (Larson 1981; Solomon et al. 1987), which is generally explained by the 
increasing contribution of turbulence at larger cloud scales (see also discussion in Heyer et al. 2009, Ballesteros-Paredes et al. 2011, and Dobbs et al. 
2014). However, the transonic properties of the Cassiopeia (east) Filament show a large departure from this expected relation. The Cassiopeia Filament, 
together with Musca, show observational evidences of filamentary molecular clouds clearly decoupled from the turbulent regime over multi-scales (from 
roughly 10\,pc to $\sim$\,100\,pc). Therefore, the Cassiopeia Filament may offers us a unique case to study not only the formation of large-scale filaments 
in the Galaxy but also the origin of non-thermal linewidths in the molecular clouds.

\subsection{The Effects from Stellar Feedback}

By ejecting energy, momentum and material into the surrounding environment, stellar feedback plays an important role in the distribution and properties 
of the ISM and also drives the recycle of the ISM and the evolution of the Galaxy. Assuming that the observed shell is resulted from a wind-blown bubble, 
we discuss the effects from stellar feedback on the Cassiopeia Filament, in addition to the broaden linewidths discussed above.

\subsubsection{The Inputted Energy}

Using the method suggested by Weaver et al. (1977), the value of the mechanical luminosity of the stellar wind ($L_{\rm wind}$) can be calculated by 
$L_{\rm wind}$\,$\approx$\,$\frac{1}{3}$$\frac{n_{\rm gas}}{{\rm cm^{-3}}}$($\frac{R_{\rm b}}{{\rm pc}}$)$^2$($\frac{V_{\rm b}}{{\rm km\,s^{-1}}}$)$^3$\,$\times$\,
10$^{30}$~erg\,s$^{-1}$, in order to excavate a bubble with a radius of $R_{\rm b}$ and expansion velocity of $V_{\rm b}$ in a gas cloud with a density of 
$n_{\rm gas}$. The density can be measured from the radius and column density by  $n_{\rm gas}$ = $\frac{3{\it N_{\rm shell}}}{{\it R_{\rm b}}}$ (see Weaver 
et al. 1977), where $N_{\rm shell}$ is the column density observed at the shell (roughly 2-3\,$\times$\,10$^{21}$\,cm$^{-2}$; see Fig.\,8c). Adopting the
radius $R_{\rm b}$ of 16\,pc, the estimated $n_{\rm gas}$ is about 150\,cm$^{-3}$. For the giant bubble inferred in this work ($V_{\rm b}$\,$\sim$\,6.5\,\kms), 
the estimated $L_{\rm wind}$ is then $\sim$\,3.5\,$\times$\,10$^{36}$\,erg\,s$^{-1}$.
The timescale of the wind needed for opening such a bubble ($\frac{16}{27}$$\frac{R_{\rm b}}{\rm pc}$$\frac{\rm km\,s^{-1}}{V_{\rm b}}$\,$\times$\,10$^6$\,yr; 
Weaver et al. 1977) is $\sim$\,1.4\,$\times$\,10$^6$\,yr. 

Adopting the gas mass of $\sim$\,6100\,$M_\odot$ ($X$-factor method) for the shell and an expanding velocity of $\sim$\,6.5\,\kms, the estimated kinetic energy 
of the arc-like shell is $\sim$\,2.3\,$\times$\,10$^{48}$\,erg. The estimated wind mechanical energy ($L_{\rm wind}$\,$\times$\,timescale) is roughly 
1.5\,$\times$\,10$^{50}$\,erg, which implies that about 2\% of the wind mechanical energy is transferred into the kinetic energy of the expanding gas shell. 

\subsubsection{The Triggered Star Formation}

Stars are formed in dense molecular cloud cores. It has been long suggested that new generation star formation can be triggered in the dense shells of 
the large bubbles produced by massive stars (e.g., Elmegreen 1998). This so-called `collect and collapse' mode has been extensively studied in the past 
decades (see, e.g., Dale et al. 2015 and references therein). Indeed, in the infrared observations, young stellar objects (YSOs), Class\,I and Class\,II objects 
(see Evans 1999; McKee \& Ostriker 2007), are frequently seen around the bubbles associated with HII regions (see, e.g., Watson et al. 2010), stellar winds 
(e.g., Cichowolski et al. 2015), and even SNRs (e.g., Billot et al. 2010; Zhou et al. 2014). However, after comparing the characteristic lifetimes of the 
YSOs\footnote{The typical lifetime of the Class\,I YSOs is $\sim$\,0.5\,$\times$\,10$^6$\,yr, while the lifetime of the Class\,II YSOs ranges from a few to several 
million years (see McKee \& Ostriker 2007 and Evans et al. 2009).} with the dynamical ages of the bubbles, the general conclusion is that the formation of those 
YSOs started before the formation of the bubbles (see discussion in Dale et al. 2015).

In the MWISP CO observations of the shell, faint C$^{18}$O\,(1--0) line emission is detected at the tips of Fingers 1\,\&\,4 (see Fig.\,6).
Compared with $^{12}$CO and $^{13}$CO, C$^{18}$O traces higher density gas. Assuming the LTE condition and an isotopic ratio 
[$^{16}$O/$^{18}$O] = 560 (Wilson \& Rood 1994), the mass of the two C$^{18}$O clumps is estimated to be $\sim$\,4.7\,$M_\odot$ (Finger~4) 
and 36.4\,$M_\odot$ (Finger~1), respectively. The virial parameter, representing the ratio between the kinetic and half gravitational potential 
energy, can be estimated as $\alpha_{\rm vir}=\frac{M_{\rm vir}}{M_{\rm LTE}}$, and the virial mass can be expressed as 
\begin{equation}\label{f.vmass}
M_{\rm vir}=\frac{5\delta_{\rm v}^{2}R}{G}\approx209(\frac{R}{\rm pc}) (\frac{\Delta v}{\rm{km~s^{-1}}})^{2} M_{\odot}\;,
\end{equation}
where $\Delta v$ is the FWHM line width. Based on the MWISP C$^{18}$O observations, the virial mass of the two clumps is estimated to be $\sim$\,3.5\,$M_\odot$ (Finger~4) 
and 16.5\,$M_\odot$ (Finger~1), leading to the $\alpha_{\rm vir}$ of $\sim$\,0.74 and $\sim$\,0.45, respectively.

In analyses, the virial parameters are used to evaluate whether objects are gravitationally bound or unbound, and $\alpha_{\rm vir}=2$ is generally
regarded as the upper limit of the critical virial parameter (see, e.g., Kauffmann et al. 2013).  The low virial parameters of the two C$^{18}$O clumps 
suggest that the two clumps are gravitationally bound. After checking infrared archival images (e.g., WISE data, not shown here), we do not find any 
infrared sources spatially coincident with the two C$^{18}$O clumps. Therefore, the two clumps are likely still in the prestellar phase, which is consistent 
with the dynamical age of the bubble/shell estimated above. We then suggest that early phase star formation has been triggered in the finger tips of 
the wind-blown bubble.
It is of interest to note that a number of prestellar and protostellar cores were also detected in the east shell of the bubble (see Figure~12) in the Herschel 
Hi-Gal survey (Elia et al. 2021). This may imply that early phase star formation is also triggered in this region. Further observations are needed to study star 
formation in the shell and analyze the effects of `collect and collapse' mode in the Cassiopeia Filament.

\subsubsection{The Rayleigh-Taylor Fingers?}

The fingers detected in the shell look like the fingers or pillars found in the numerical simulations  (e.g., Freyer et al. 2003; Wareing et al. 2017) and 
observations (e.g., Deharveng et al. 2012; Schneider et al. 2016) in the studies of massive stellar feedback. Nevertheless, the observed fingers in this 
work show two interesting characteristics: (1) all the fingers protrude outward, different from those fingers/pillars found in the massive star-forming
regions (which protrude inward toward driving stars or clusters), and (2) the shell is not spatially associated with ionized gas (HII regions).

As we discussed in $\S$\,4.2, the expanding arc-like shell cloud be produced by the stellar wind of the progenitor star of an SNR. This may explain
the non-association between the shell (stellar wind feedback) and ionized gas (stellar ionization feedback). Interestingly, in the numerical studies (see, 
e.g., Chevalier et al. 1992; Chevalier \& Blondin 1995; Blondin et al. 1996; Tutone et al. 2020), finger-like structures, due to the Rayleigh-Taylor (RT) 
instability, are seen in the circumstellar shells around supernovae. These RT fingers, all protruding outward, are the results from the interaction between 
the supernova ejecta and shocked shell. In the observations, the RT fingers are used to explain the protrusions frequently detected around the peripheries 
of the SNRs, such as Cassiopeia A (Orlando et al. 2021) and Vela (Wang \& Chevalier 2002; Miceli et al. 2013). The fingers found in this work are likely 
also the RT fingers, given that the shell was produced by the stellar wind of a progenitor star.

\section{SUMMARY}

We present large-field CO\,(1--0) molecular line observations toward the Galactic Cassiopeia region, using the PMO 13.7\,m telescope. The main
results of this work are summarized below.

(1) The CO observations reveal  a large-scale highly filamentary molecular cloud within the region of 132\fdg0\,$\geq$\,$l$\,$\geq$\,122\fdg0 and 
$-$1\fdg0\,$\leq$\,$b$\,$\leq$\,3\fdg0 and the velocity range from approximately +1 to $+$4\,\kms. Based on the CO data, we derive the excitation
temperature, column density, and mass of the filamentary molecular cloud.

(2) The measured length of the main filament, named after the Cassiopeia Filament, is $\sim$\,390\,pc, making it one of the longest molecular 
filaments in the Galaxy. The observed properties of the Cassiopeia Filament are similar to those synthetic large-scale filament in the inter-arm 
regions, as well as the large-scale filaments observed in the outer Galaxy inter-arm regions. Based on its observed properties and location on 
the Galactic plane, we suggest that the Cassiopeia Filament is a spur of the Local arm, which is formed due to the galactic shear.

(3) A giant arc-like molecular gas shell is found in the west of the Cassiopeia Filament, which covers an extent with coordinates of 
125\fdg5\,$\geq$\,$l$\,$\geq$\,121\fdg5 and $-$1\fdg3\,$\leq$\,$b$\,$\leq$\,0\fdg5, and a velocity range between $-$1 and +7\,\vkm. 
Finger-like structures, with systematic velocity gradients, are detected in the shell. The observed CO gas kinematics suggest that the 
large shell is expanding at a velocity of $\sim$\,6.5\,\kms. 

(4) Both the shell and finger-like structures outline a giant bubble with a radius of $\sim$\,16\,pc, centered at $l$\,=\,124\fdg06, $b$\,=\,1\fdg15. 
We suggest that the bubble is produced by stellar wind, which is likely from the progenitor star of an SNR in the region. Nevertheless, further 
observations are needed to verify the candidate SNR suggested in this work. 

(5) The observed CO linewidths in the west shell are much larger those found in the east filament, implying that strong turbulence is driven by 
stellar wind into the shell. Based on the observations, we suggest that the whole Cassiopeia Filament could be quiescent initially until its west 
part was blown by stellar wind. Without external disturbances, the east filament is still transonic in the $^{12}$CO observations and subsonic 
in the $^{13}$CO observations.

\acknowledgments

We thank the anonymous referee for providing insightful suggestions and comments, which help us to improve this work.
This work was supported by the National Natural Science Foundation of China (grant No. 12041305), the National Key R\&D Program of 
China (grant No. 2017YFA0402702), and the CAS International Cooperation Program (grant No. 114332KYSB20190009). 
We thank Mr. Chaojie Hao for helping us to plot the projected Galactic plane image.
We are grateful to all the members of the Milky Way Imaging Scroll Painting (MWISP) CO line survey group, especially the staff of Qinghai
Radio Station of PMO at Delingha for the support during the observations. MWISP is sponsored by the National Key R\&D Program of China 
with grant 2017YFA0402701 and the CAS Key Research Program of Frontier Sciences with grant QYZDJ-SSW-SLH047.

\clearpage

\clearpage


\begin{figure*}
\begin{center}
\includegraphics[width=17cm,angle=0]{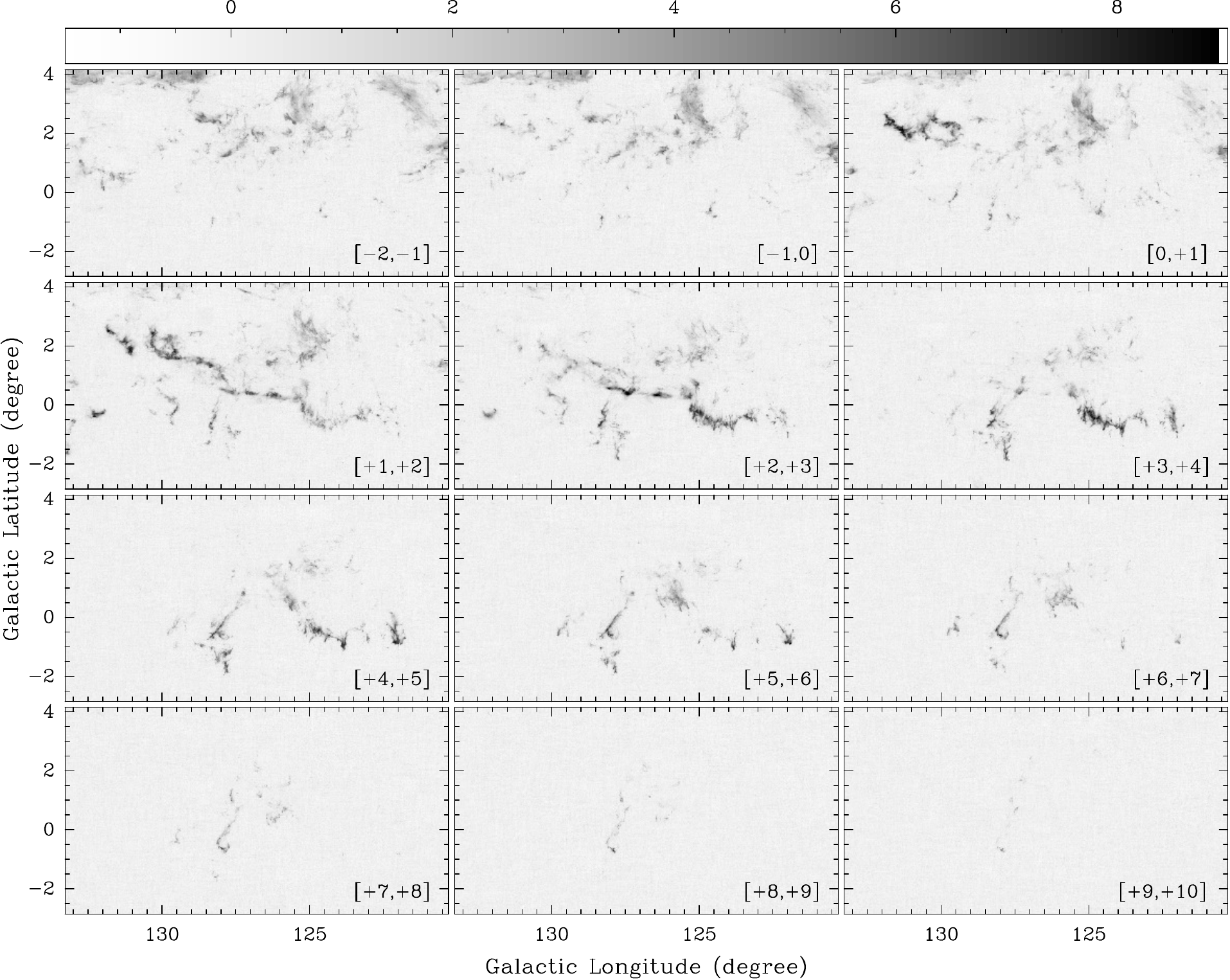}
\end{center}
\caption{The velocity-integrated intensity channel maps of the $^{12}$CO\,(1--0) emission toward the Galactic Cassiopeia region. The integrated 
velocity range is written in the bottom right corner of each panel (in km\,s$^{-1}$). The unit of the scale bar is K\,\,km\,s$^{-1}$.\label{MWISP_12CO_channel}}
\end{figure*}

\begin{figure*}
\begin{center}
\includegraphics[width=17cm,angle=0]{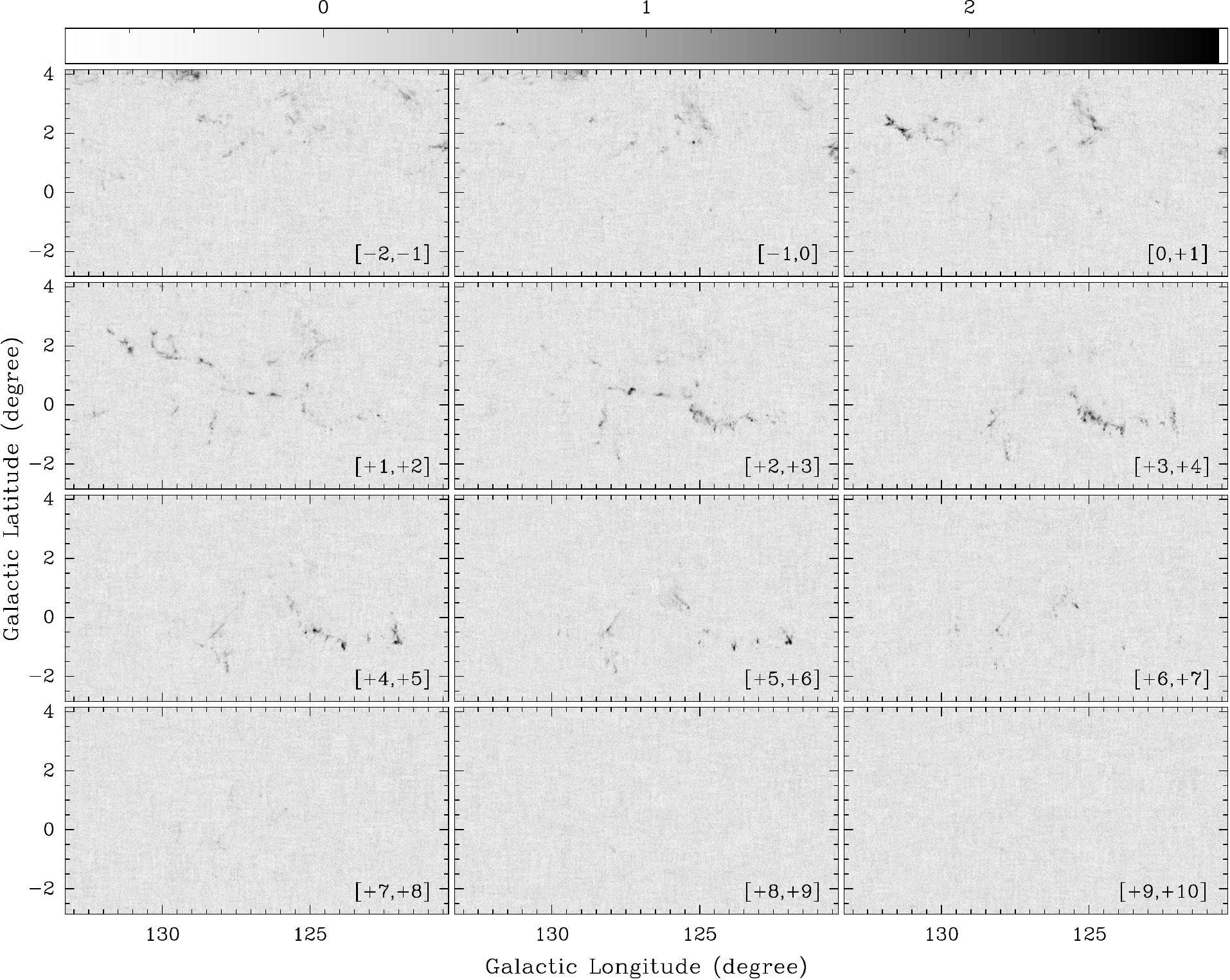}
\end{center}
\caption{The velocity-integrated intensity channel maps of the $^{13}$CO\,(1--0) emission toward the Galactic Cassiopeia region. The integrated 
velocity range is written in the bottom right corner of each panel (in km\,s$^{-1}$). The unit of the scale bar is K\,\,km\,s$^{-1}$.\label{MWISP_13CO_channel}}
\end{figure*}

\begin{figure*}
\begin{center}
\includegraphics[width=17cm,angle=0]{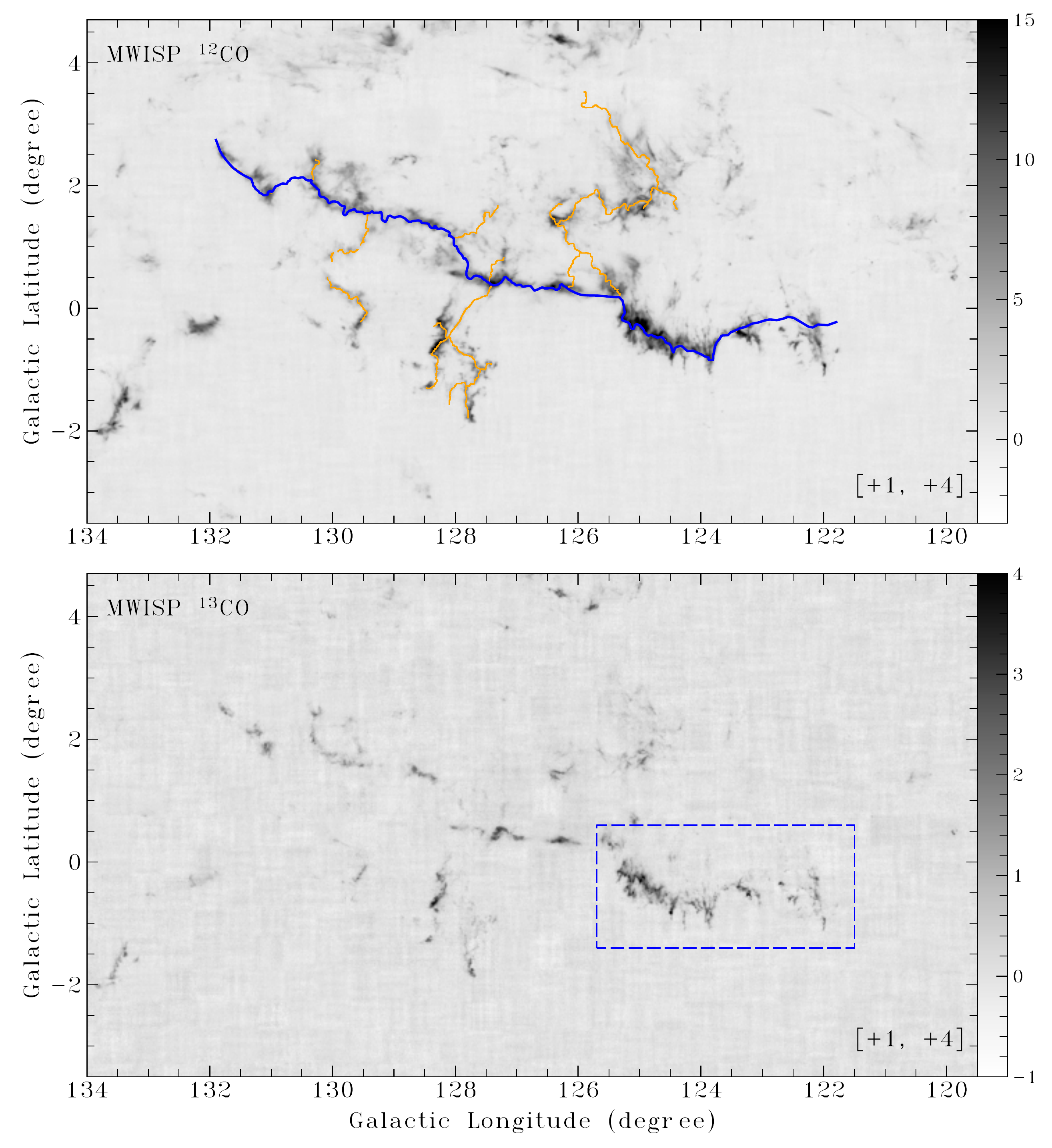}
\end{center}
\caption{The velocity-integrated intensity images of the $^{12}$CO (top panel) and $^{13}$CO (bottom panel) emission toward the Galactic Cassiopeia region.
The unit of the scale bar is K\,\,km\,s$^{-1}$. In the $^{12}$CO image, the blue line shows the main filament identified by the DisPerSE algorithm (named after 
the Cassiopeia Filament in this work), while the orange lines show the identified sub-filaments. In the $^{13}$CO image, the blue dashed-line rectangle marks 
the extent of the giant arc-like shell.\label{MWISP_CO_intensity}}
\end{figure*}

\begin{figure*}
\begin{center}
\includegraphics[width=17cm,angle=0]{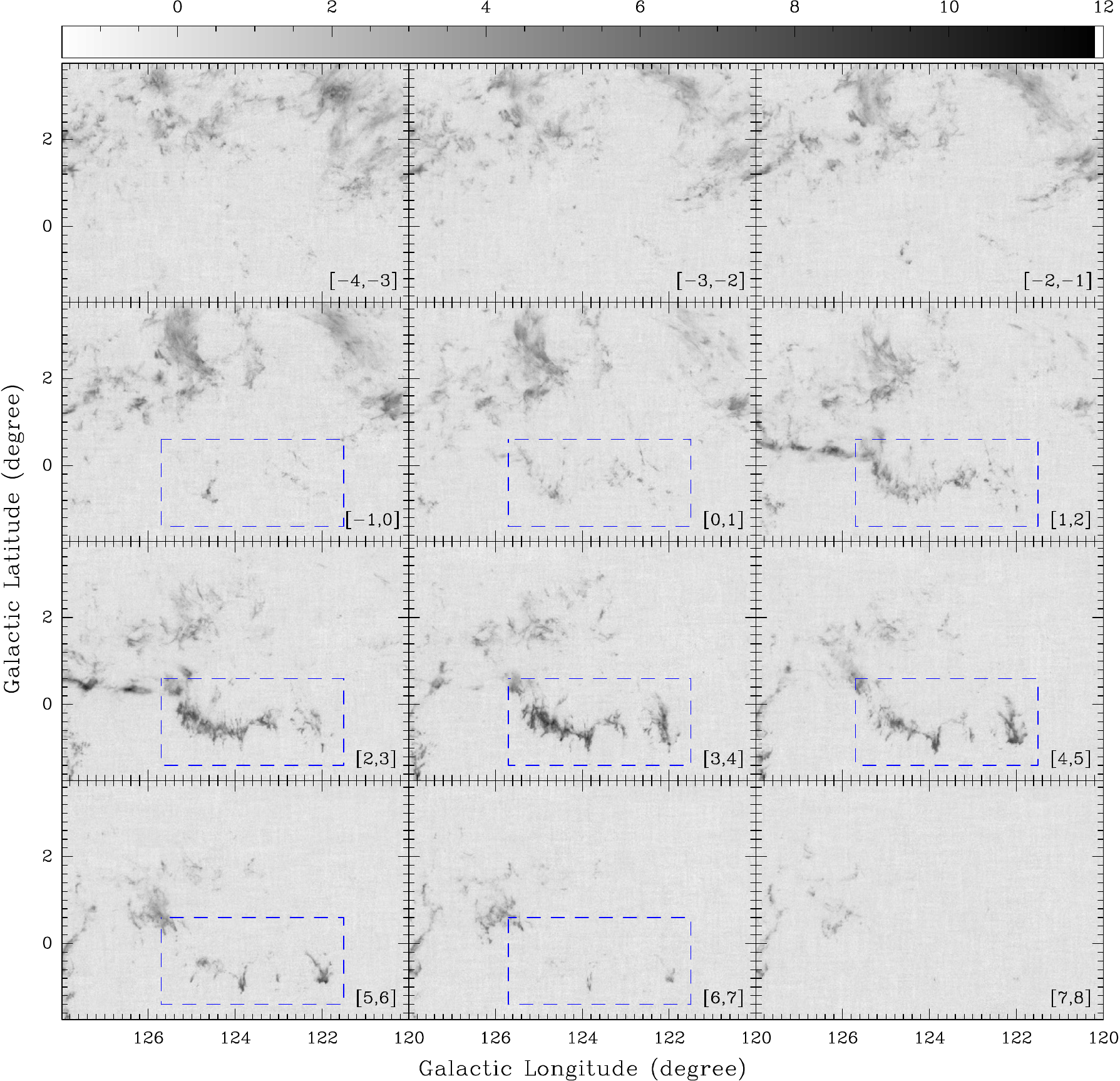}
\end{center}
\caption{The velocity-integrated intensity channel maps of the $^{12}$CO\,(1--0) emission toward the west of the Cassiopeia Filament. The integrated velocity 
range is written in the bottom right corner of each panel (in km\,s$^{-1}$). The blue dashed-line rectangle in the panels shows the extent of the arc-like shell 
found in the velocity range from roughly $-$1 to +7\,\kms.
\label{MWISP_shell_channel}}
\end{figure*}

\begin{figure*}
\begin{center}
\includegraphics[width=17cm,angle=0]{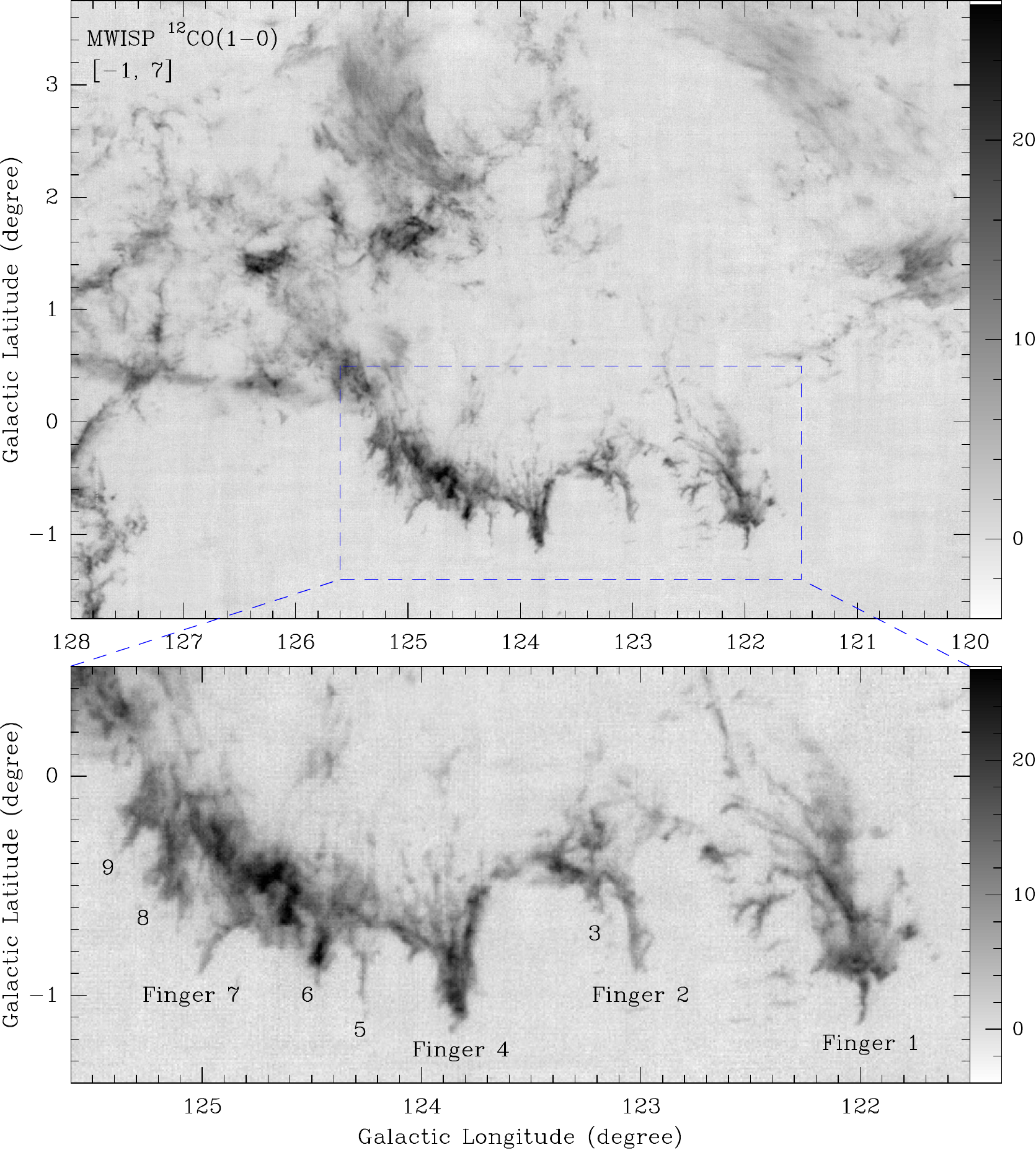}
\end{center}
\caption{{\it Top:} The $^{12}$CO\,(1--0) velocity-integrated intensity image of the shell region. The unit of the scale bar is K\,km\,s$^{-1}$. The $^{12}$CO emission 
is integrated between $-$1 and +7\,km\,s$^{-1}$ (1\,$\sigma$\,$\sim$\,0.8\,K\,km\,s$^{-1}$). 
{\it Bottom:} The enlarged view of the shell and finger-like structures.
\label{MWISP_12CO_shell}}
\end{figure*}

\begin{figure*}
\begin{center}
\includegraphics[width=16cm,angle=0]{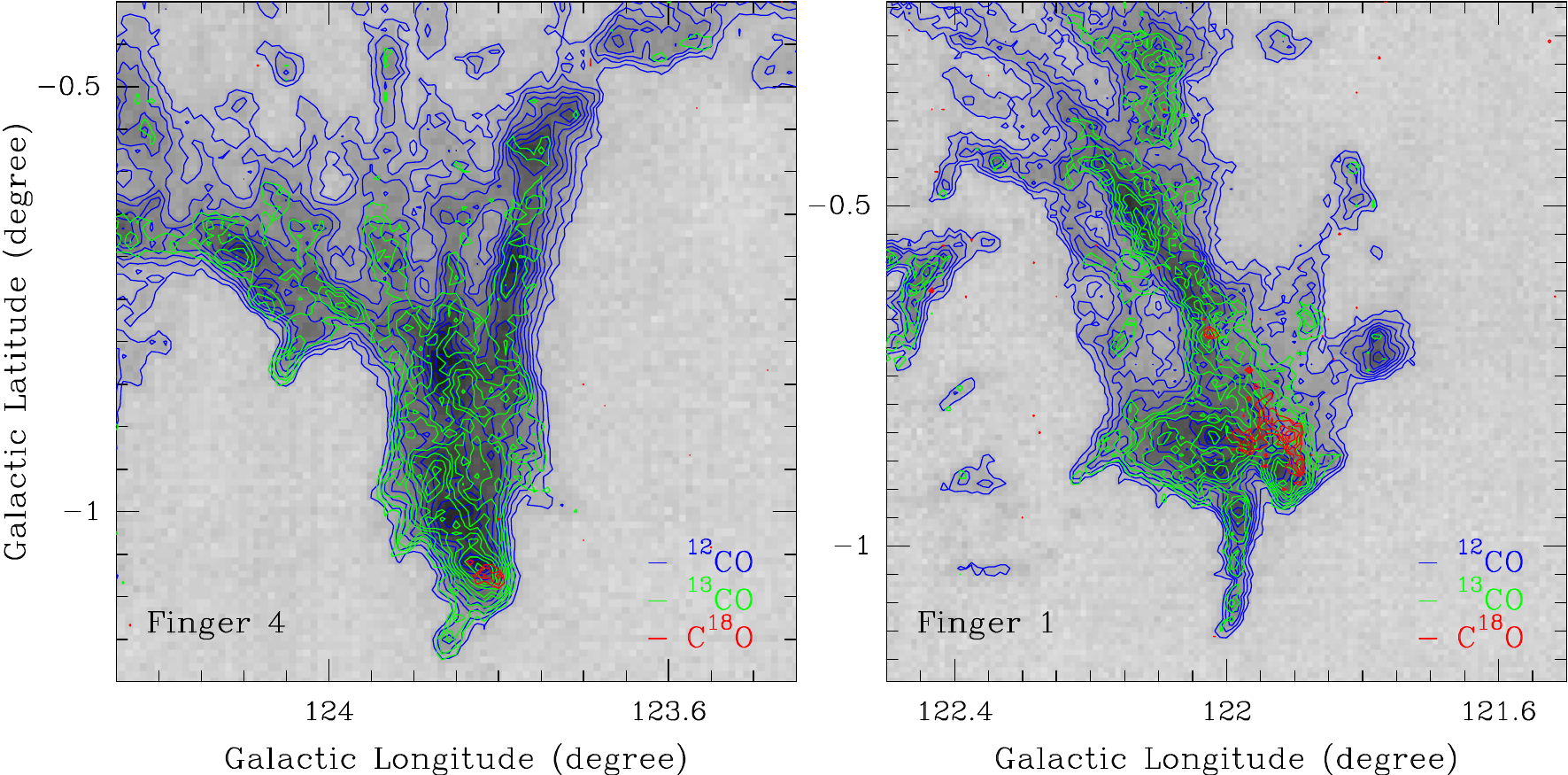}
\end{center}
\caption{The CO\,(1--0) multi-line emission of Finger~4 ({\it left}) and Finger~1 ({\it right}). In the two panels, the $^{12}$CO emission contours (blue; integrated 
between $-1$ and +7\,\kms) start at 5\,$\sigma$ and increase in steps of 3\,$\sigma$ (1\,$\sigma$\,$\sim$\,0.8\,K\,km\,s$^{-1}$). The $^{13}$CO emission contours 
(green; integrated between 0.5 and 6.5\,\kms) start at 4\,$\sigma$ and increase in steps of 2\,$\sigma$ (1\,$\sigma$\,$\sim$\,0.37\,K\,km\,s$^{-1}$). The C$^{18}$O 
emission integrated between 4.5 and 6.5\,km\,s$^{-1}$, and the contours (red) start at 3\,$\sigma$ and increase in steps of 1\,$\sigma$ (1\,$\sigma$\,$\sim$\,0.21\,K\,km\,s$^{-1}$).\label{MWISP_C18O}}
\end{figure*}

\begin{figure*}
\gridline{\fig{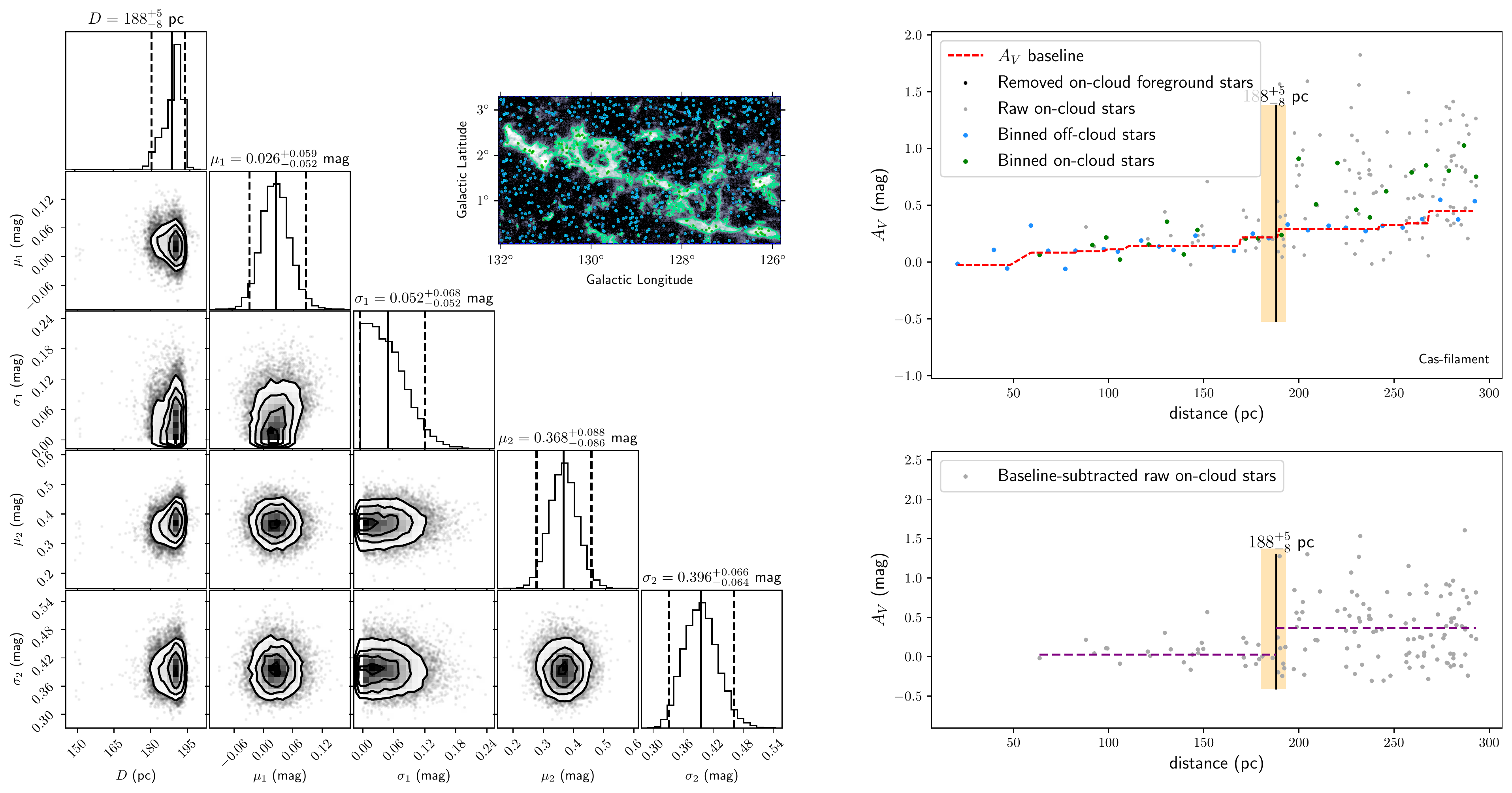}{0.75\textwidth}{(a)}}
\gridline{\fig{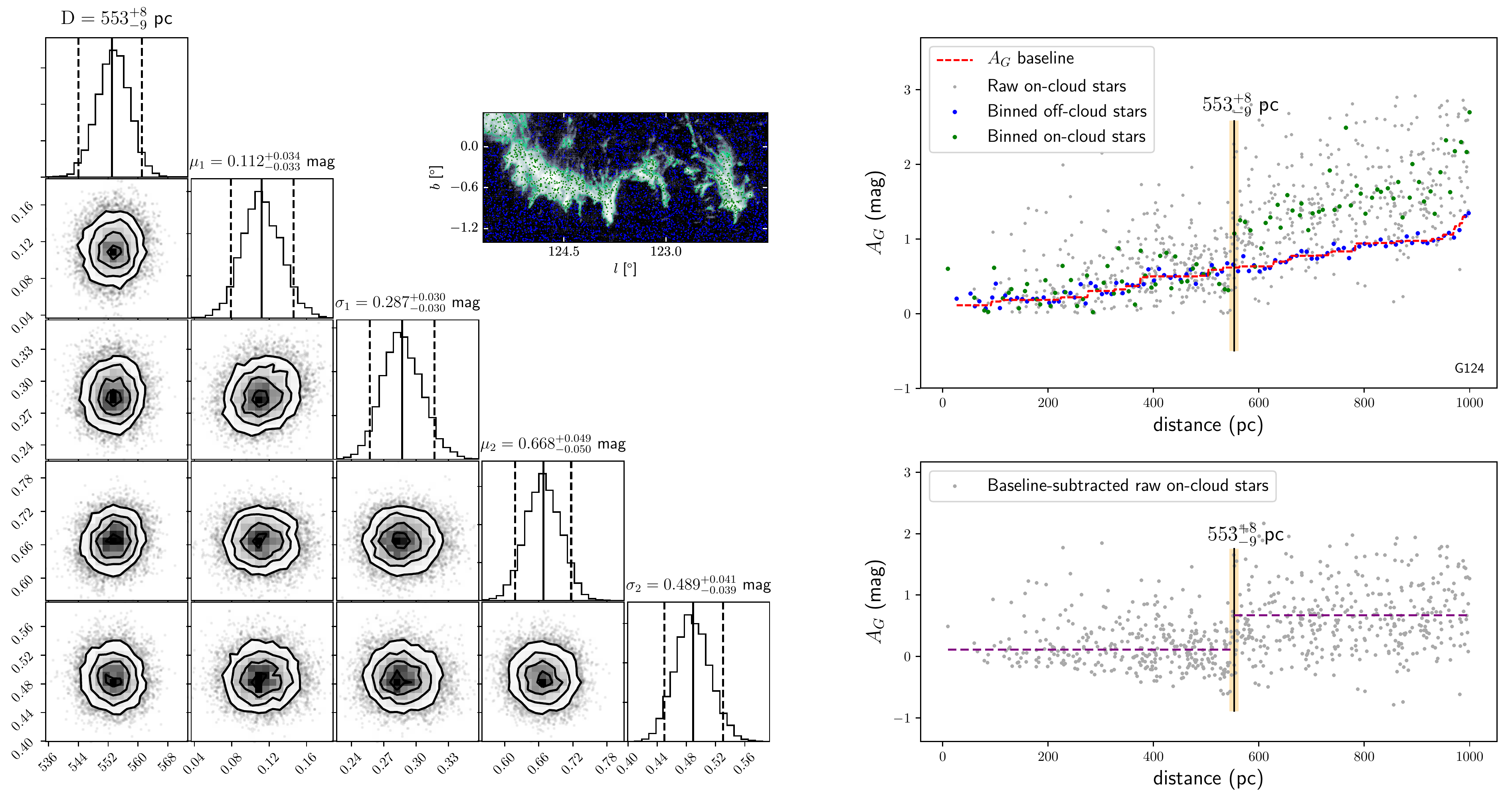}{0.75\textwidth}{(b)}}
\caption{\footnotesize (a) The measured distance toward the filament east end. The middle panel shows the MWISP $^{12}$CO intensity image. 
The green contour shows the edge of molecular cloud (3\,$\sigma$ threshold), while green and blue dots represent on- and off-cloud $Gaia$ DR2 
stars. In the top right panel, green and blue points represent on- and off-cloud stars (binned every 5\,pc), respectively. The dashed red line is the 
modeled extinction $A_{V}$. The distance is derived with raw on-cloud $Gaia$ DR2 stars, which are represented with gray points. The black vertical 
lines indicate the distance ($D$) estimated with Bayesian analyses and Markov Chain Monte Carlo (MCMC) sampling, and the shadow area depicts 
the 95\% highest posterior density (HPD) distance range. The corner plots of the MCMC samples are displayed in the left panel (distance, the extinction 
of foreground and background stars and uncertainties). The mean and 95\% HPD of the samples are shown with solid and dashed vertical lines, 
respectively. (b) The measured distance toward the west shell, and $Gaia$ $G$-band data are used.
\label{distance}}
\end{figure*}

\begin{figure*}
\begin{center}
\includegraphics[width=13cm,angle=0]{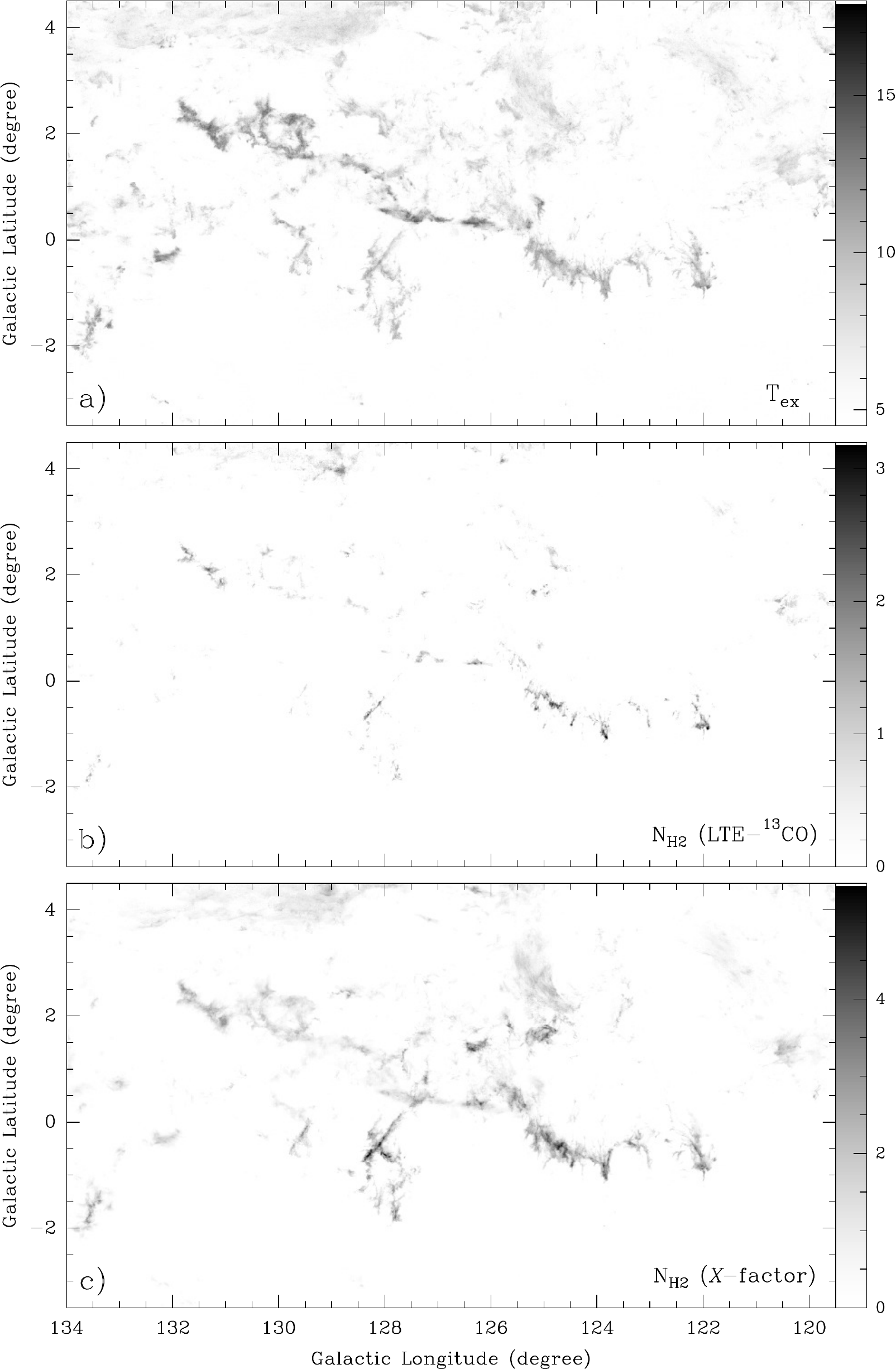}
\end{center}
\caption{(a) The excitation temperature image of the observed Cassiopeia region. The unit of the scale bar is K. (b) The H$_2$ column density image 
derived from the $^{13}$ CO data within the velocity range of [+1, +4]\,\kms, assuming the LTE condition. The unit of the scale bar is 10$^{21}$\,cm$^{-2}$. 
(c) The H$_2$ column density image derived from the $^{12}$ CO data within the velocity range of [+1, +4]\,\kms, adopting the CO-to-H$_2$ conversion 
factor. The unit of the scale bar is 10$^{21}$\,cm$^{-2}$.\label{MWISP_CO_properties}}
\end{figure*}

\begin{figure*}
\begin{center}
\includegraphics[width=16cm,angle=0]{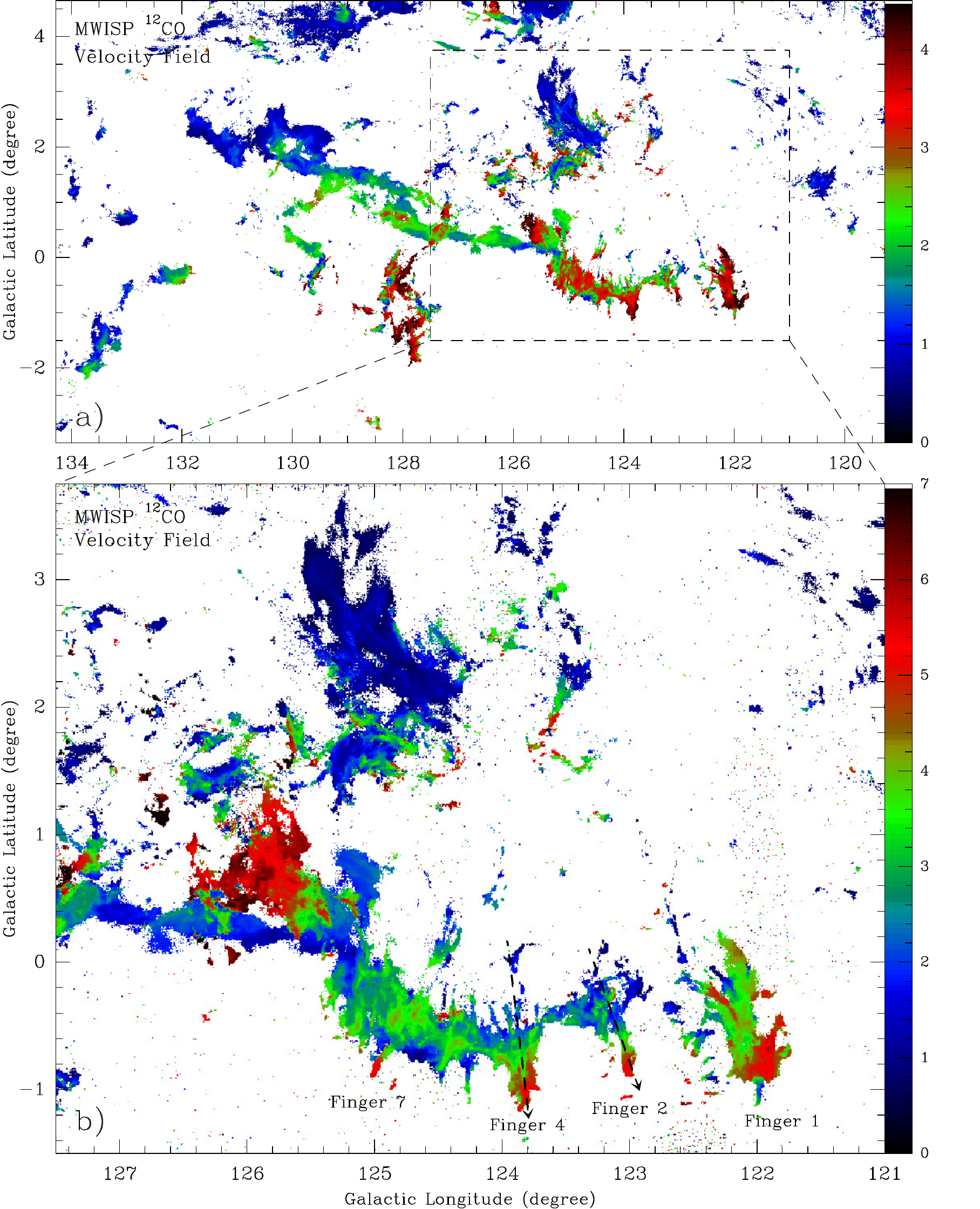}
\end{center}
\caption{{\it Top panel:} The $^{12}$CO velocity field (moment 1 image) of the observed Cassiopeia region within the velocity range of [0.5, 4.5]\,\vkm.
The unit of the scale bar is \kms. {\it Bottom panel:}  The $^{12}$CO velocity field of the shell region within the velocity range of [0, 7]\,\vkm. The black 
arrow lines show the routings of the PV diagram of the Fingers 2 and 4.\label{MWISP_CO_velfield}}
\end{figure*}

\begin{figure*}
\begin{center}
\includegraphics[width=7cm,angle=-90]{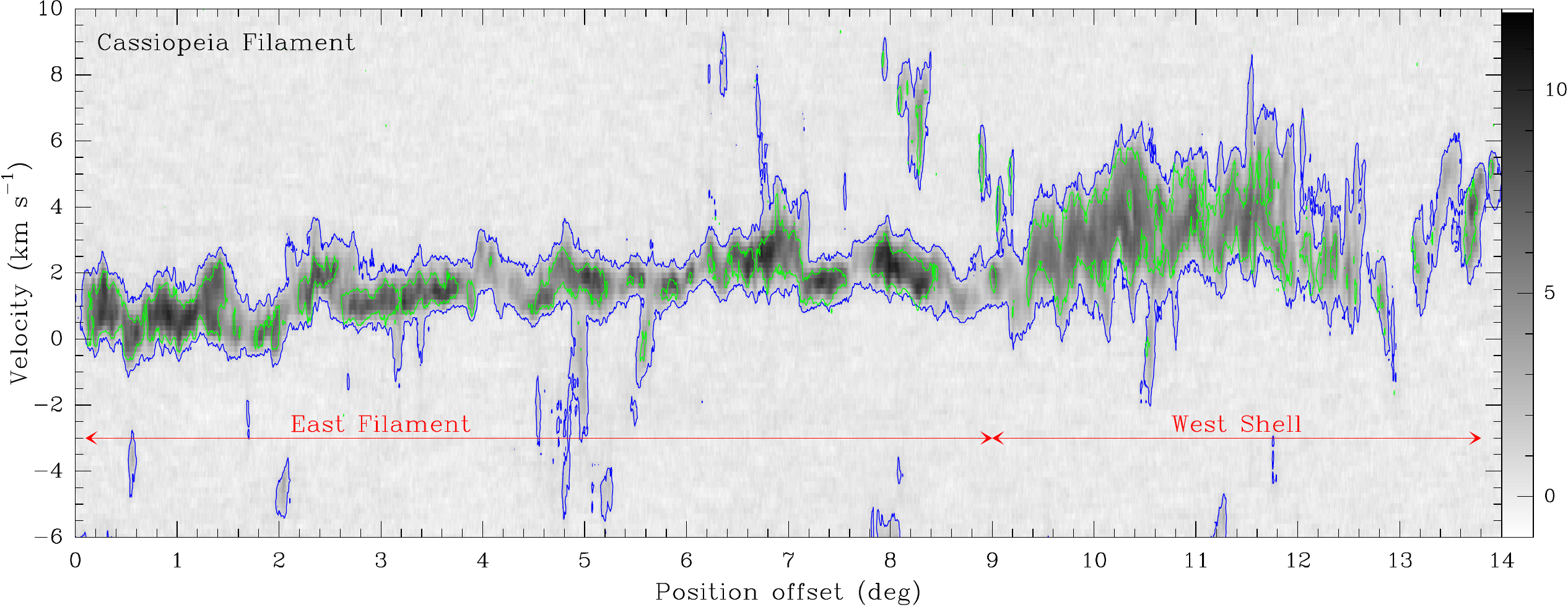}
\end{center}
\caption{The CO position-velocity (PV) diagram along the Cassiopeia Filament (from east to west). The grey-scale background image represents the $^{12}$CO 
emission, with the 5\,$\sigma$ contour (blue; 1\,$\sigma$\,$\sim$\,0.2\,K). The green contour shows the 4\,$\sigma$ $^{13}$CO emission 
(1\,$\sigma$\,$\sim$\,0.12\,K).\label{MWISP_PV_filament}}
\end{figure*}

\begin{figure*}
\begin{center}
\includegraphics[width=13cm,angle=0]{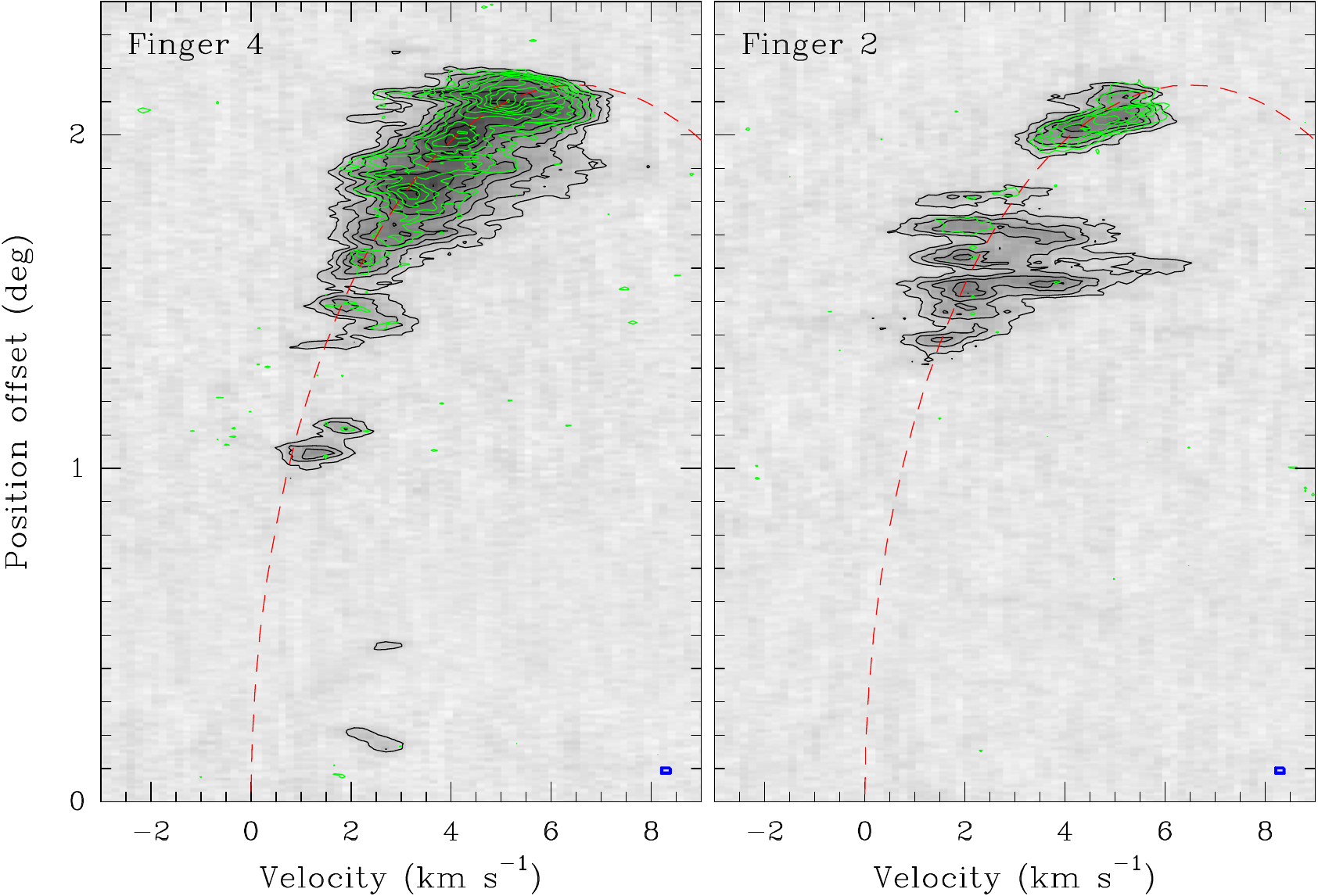}
\end{center}
\caption{The CO PV diagrams along Fingers 4 ({\it left}) and 2 ({\it right}). The grey-scale background image represents the $^{12}$CO emission. The $^{12}$CO 
contours (black) start from 1\,K and increase by a step of 1\,K (1\,$\sigma$\,$\sim$\,0.2\,K). The $^{13}$CO emission contours (green) start from 0.4\,K and increase
by a step of 0.4\,K (1\,$\sigma$\,$\sim$\,0.12\,K). The red ellipse shows the fitting toward the PV diagram with a velocity radius of 6.5\,\kms.\label{MWISP_PV_finger}}
\end{figure*}

\begin{figure*}
\begin{center}
\includegraphics[width=16cm,angle=0]{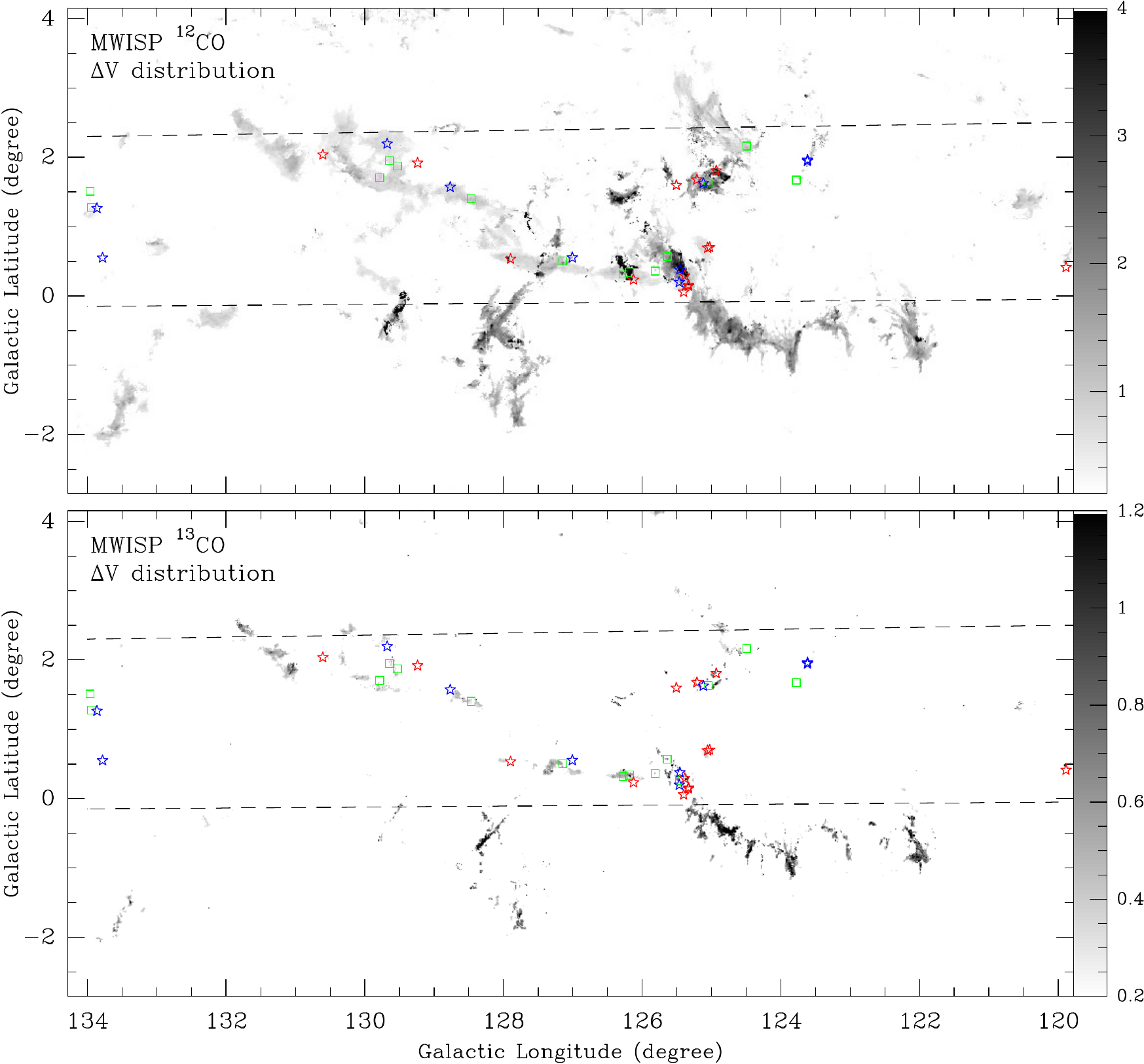}
\end{center}
\caption{{\it Top panel:} The distribution of the $^{12}$CO linewidths (moment 2 image) of the observed Cassiopeia region within the velocity range of [0, 7]\,\kms. 
The unit of the scale bar is \kms. {\it Bottom panel:} The same as the top panel, but for the $^{13}$CO emission. In the two panels, green squares, red, and blue 
stars show the positions of starless, prestellar, and protostellar cores detected in the Herschel Hi-Gal survey (see Elia et al. 2021), while the two dashed lines show 
the covered extent in the Hi-Gal survey. The Cassiopeia filament is observed in the velocity range of [+1, +4]\,\kms, the Herschel Hi-Gal cores are then selected with 
molecular gas velocities within [--1, +6]\,\kms.
\label{MWISP_CO_linewidth}}
\end{figure*}

\begin{figure*}
\begin{center}
\includegraphics[width=16cm,angle=0]{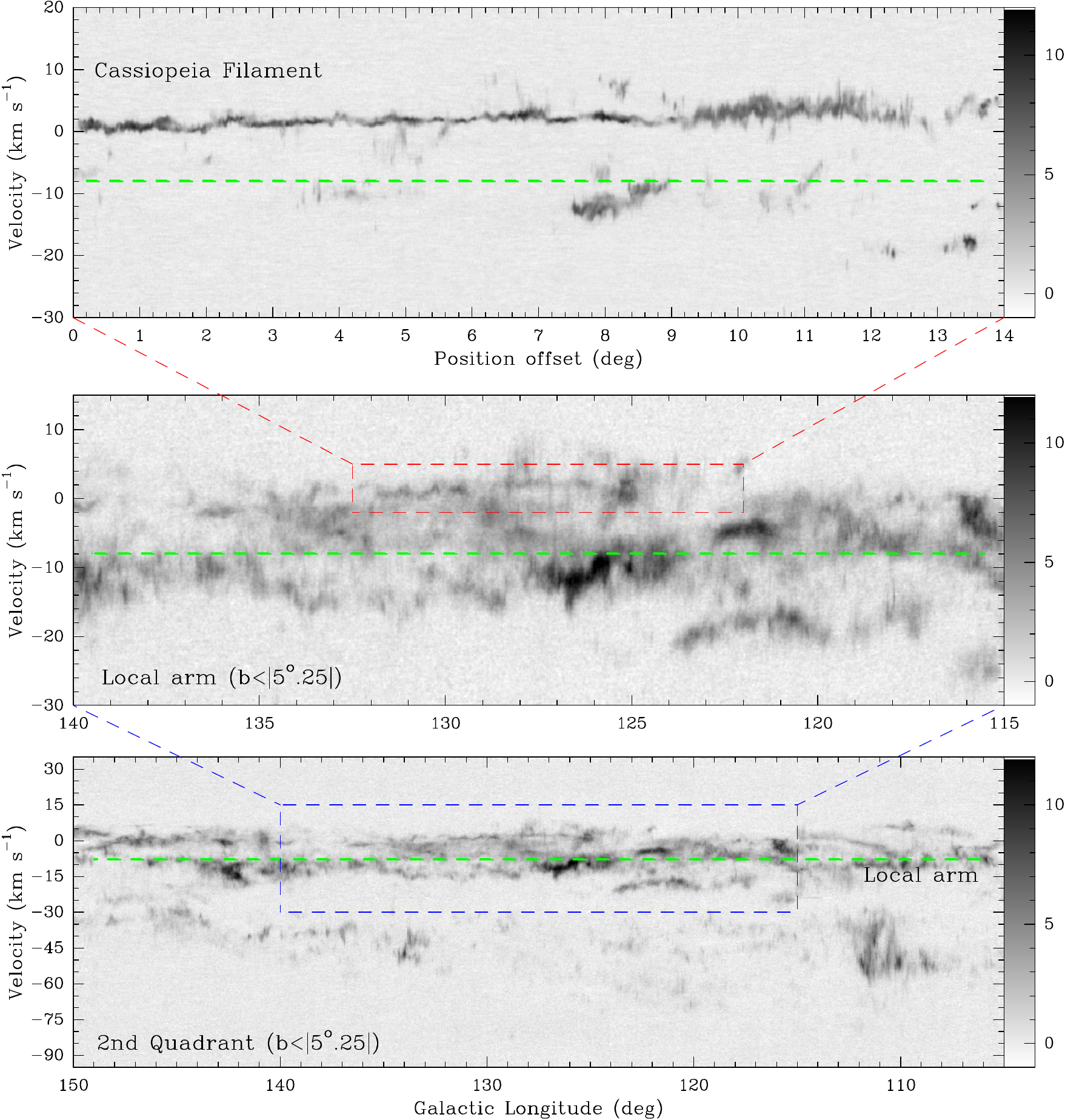}
\end{center}
\caption{{\it Bottom panel:} The CO longitude-velocity (LV) diagram for the 2nd quadrant with $b$\,$\leq$\,5\fdg25 (see also Yan et al. 2021). The green
dashed line shows the central mean velocity of the Local arm ($\sim$\,8\,\kms).  {\it Middle panel:} The enlarged view of the LV diagram for the Local
arm. The red dashed box shows the longitude and velocity range of the Cassiopeia Filament (but integrated with $b$\,$\leq$\,5\fdg25). {\it Top panel:} 
The CO PV diagram along the Cassiopeia Filament (similar to Figure~10), with respect to the central mean velocity of the Local arm.
\label{PV_new}}
\end{figure*}

\begin{figure*}
\begin{center}
\includegraphics[width=16cm,angle=0]{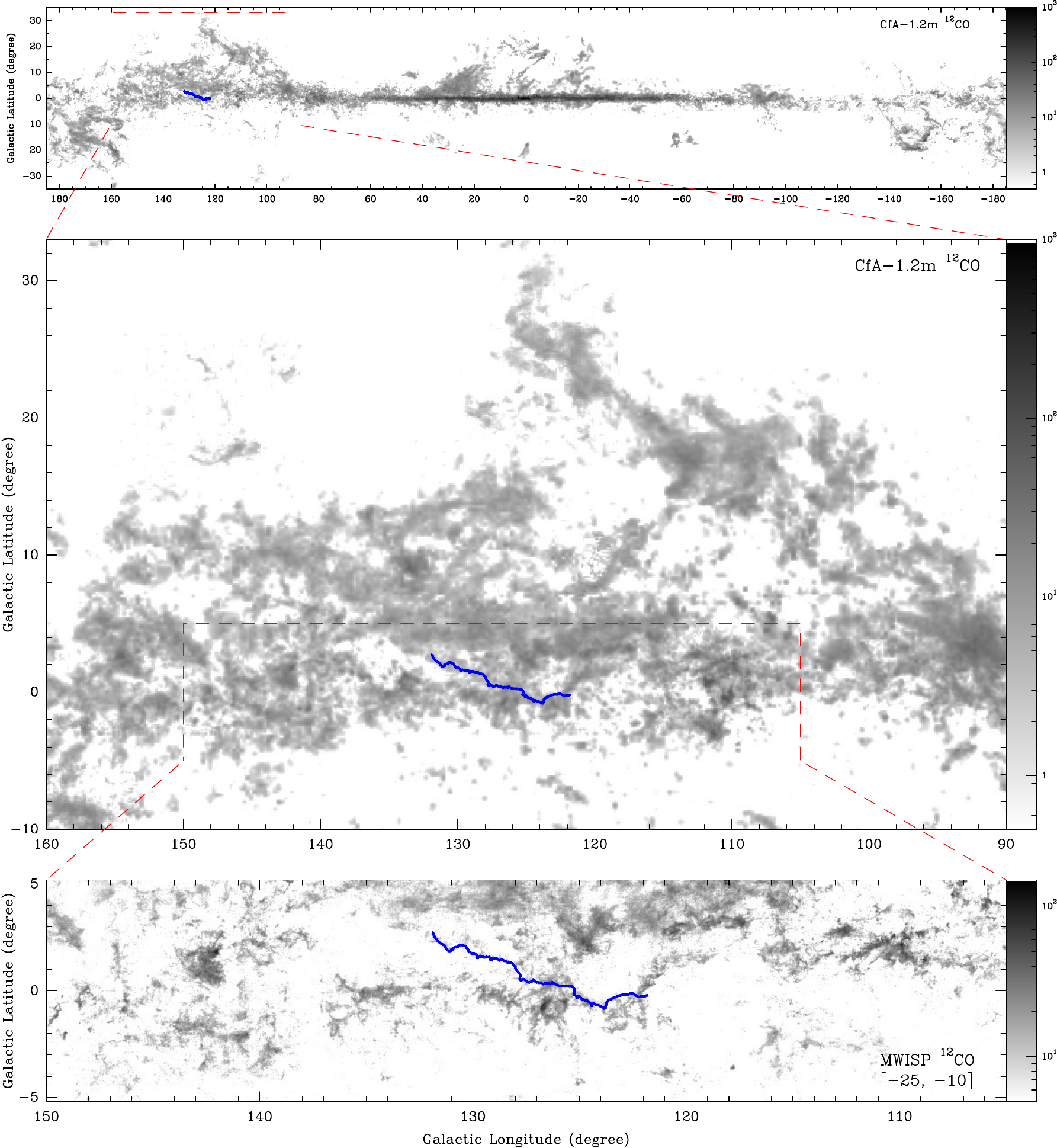}
\end{center}
\caption{{\it Top panel:} The velocity-integrated $^{12}$CO intensity map of the Milky Way obtained from the CfA-1.2\,m survey (Figure~2 in Dame et al. 2001).
{\it Middle panel:} Also the CfA-1.2\,m $^{12}$CO image, but the enlarged view toward the 2nd quadrant (including the Cassiopeia region). {\it Bottom panel:} 
The MWISP $^{12}$CO image, integrated between --25 and +10\,\kms\ (i.e., Local arm only). For each panel, the blue line shows the Cassiopeia Filament. 
Note that all CO images in this work are from the MWISP survey and only the two panels here are from the CfA 1.2\,m survey (to provide a complete molecular
gas view of the Milky Way).
\label{disk}}
\end{figure*}

\begin{figure*}
\gridline{\fig{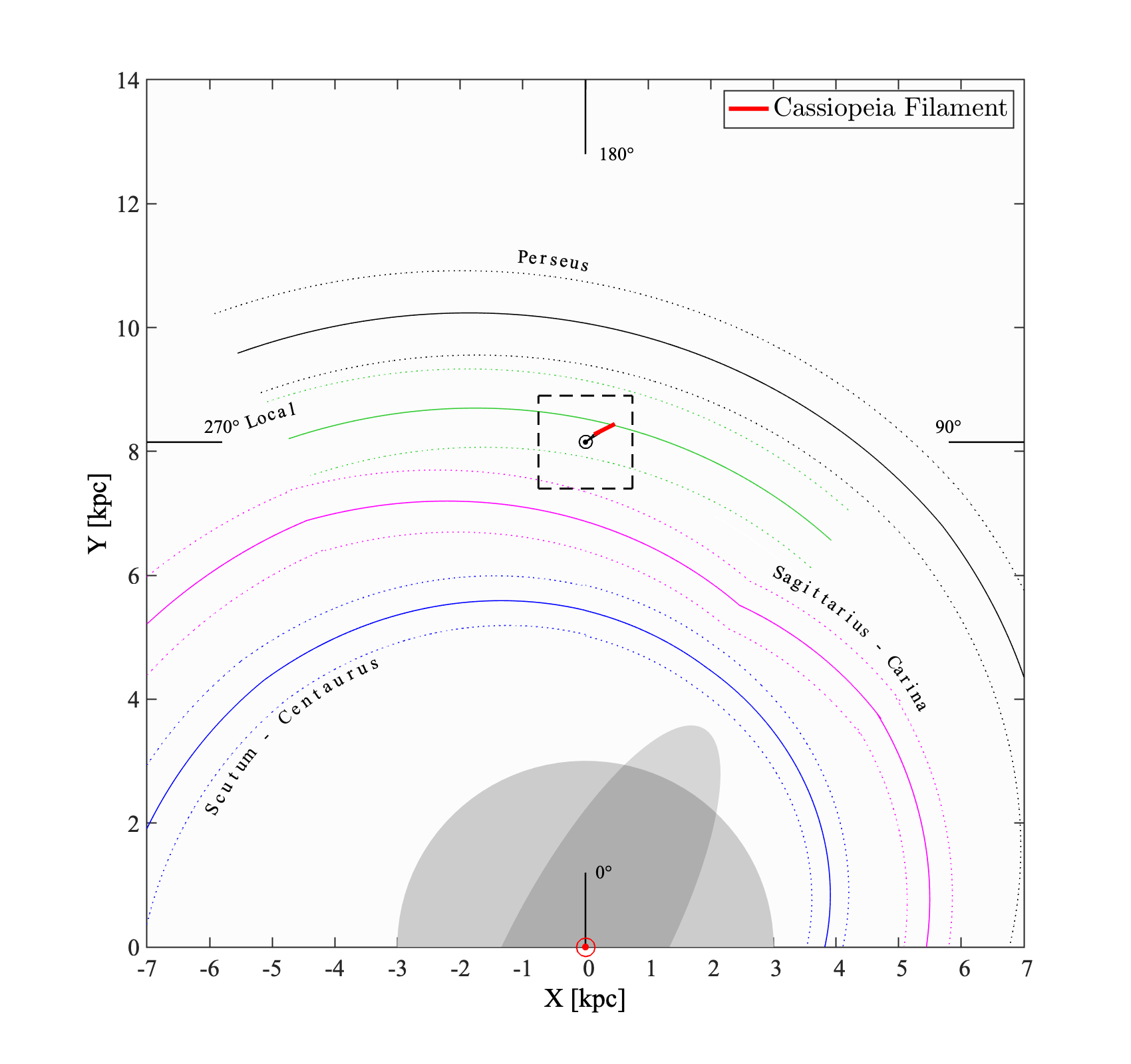}{0.5\textwidth}{(a)}
\fig{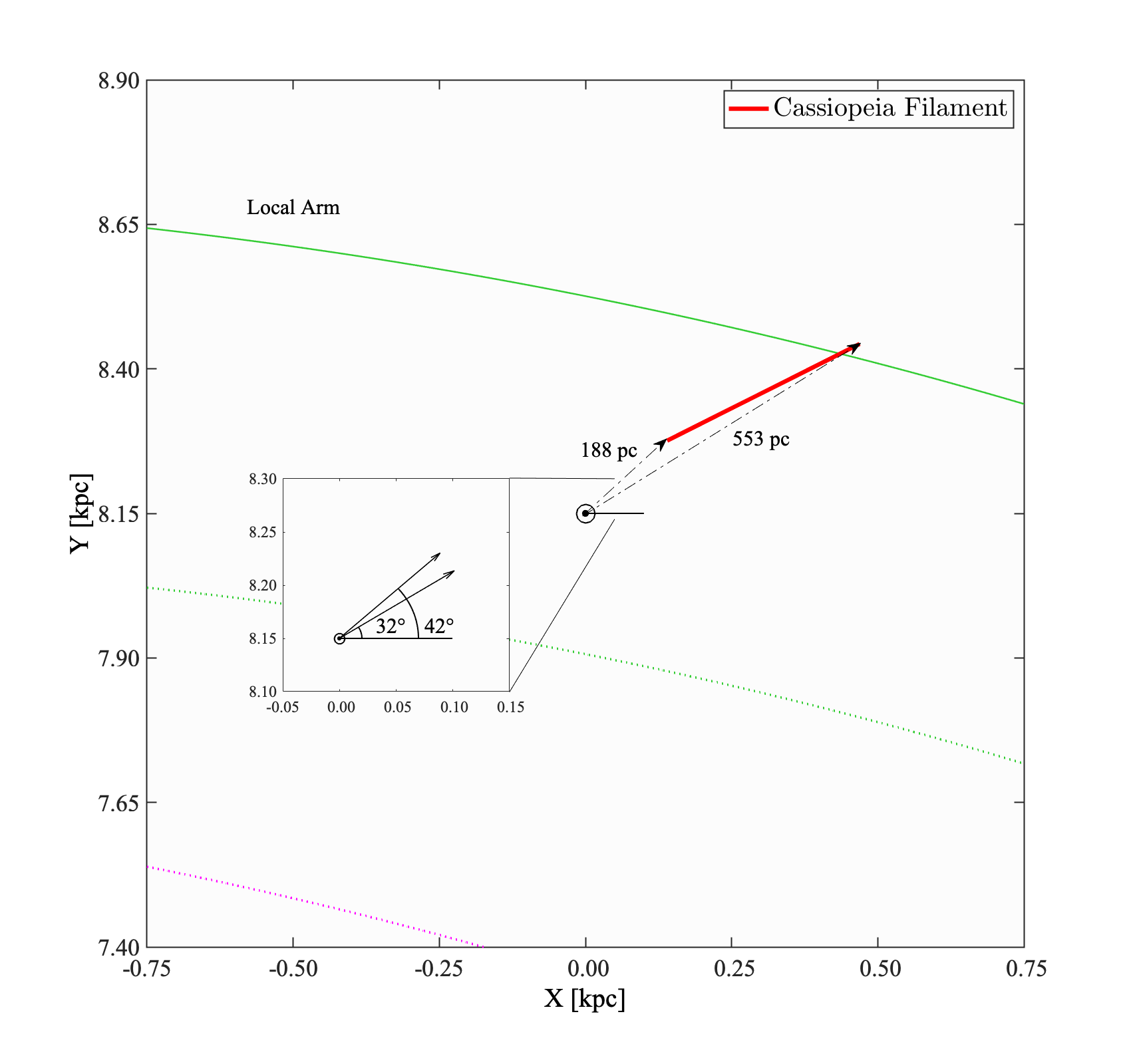}{0.5\textwidth}{(b)}}
\caption{\footnotesize (a) The locations of the Cassiopeia Filament projected onto the Galactic plane. The Galaxy center (red solar symbol) is at (0, 0) kpc, while 
the Sun (black solar symbol) is at (0, 8.15) kpc. The solid and dashed curved lines denote the arm center and widths fitted by Reid et al. (2019). (b) The enlarged 
view of the area enclosed by the dashed square in the left panel. The viewing angle of the Cassiopeia Filament is roughly 10$^\circ$ (132\fdg0\,$\geq$\,$l$\,$\geq$\,122\fdg0).
The distance of the east filament and west shell is measured to be 188\,pc and 553\,pc, respectively.
\label{spur}}
\end{figure*}

\begin{figure*}
\begin{center}
\includegraphics[width=16cm,angle=0]{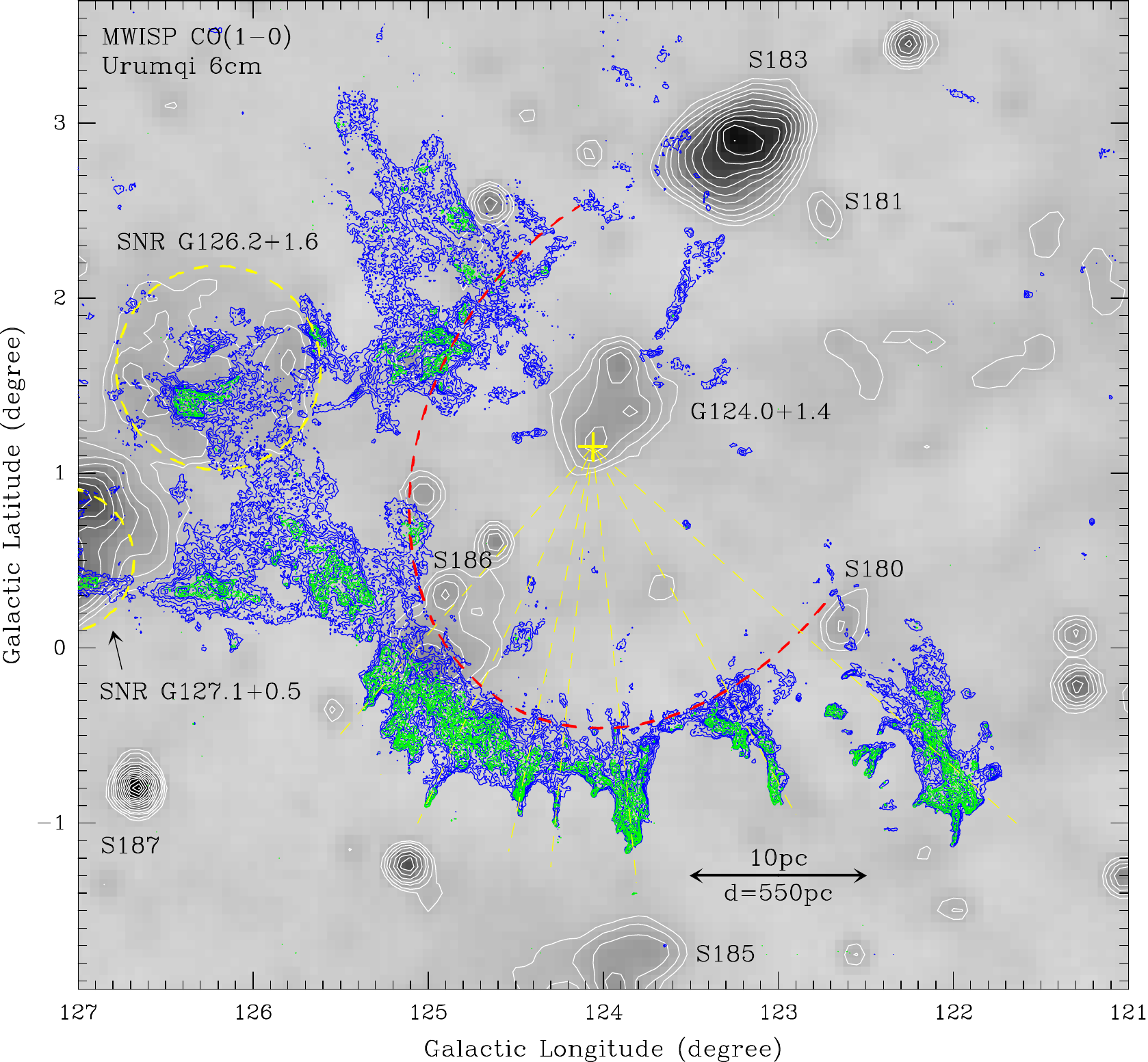}
\end{center}
\caption{The $^{12}$CO (integrated between $-$1 and +7\,\kms) and $^{13}$CO (integrated between 0.5 and 6.5\,\kms) emission contours of the 
observed Cassiopeia region, plotted on the 6\,cm radio continuum image (from Sun et al. 2007). The $^{12}$CO contours (blue) start at 5\,$\sigma$ 
and increase in a step of 3\,$\sigma$ (1\,$\sigma$ noise is $\sim$\,0.7-0.8\,K\,\kms), while the $^{13}$CO contours (green) start at 4\,$\sigma$ and 
increase in a step of 2\,$\sigma$ (1\,$\sigma$ noise is $\sim$\,0.37\,K\,\kms). The 6\,cm contours (white) correspond to 10, 15, 22, 29\,mK, and then 
increase by steps of 12\,mK (1\,$\sigma$ is $\sim$\,1.4\,mK). The red dashed partial ellipse with the radii of 3\fdg4\,$\times$\,2\fdg8 shows a potential 
bubble suggested by the morphology and kinematics of both the shell and finger-like structures, while the yellow cross shows the bubble center 
suggested by the finger-like structures ($l$\,=\,124\fdg06, $b$\,=\,1\fdg15). The yellow dashed circles show the two SNRs in the region. The names
of known HII regions are also labeled.\label{MWISP_bubble}}
\end{figure*}

\begin{figure*}
\begin{center}
\includegraphics[width=14cm,angle=0]{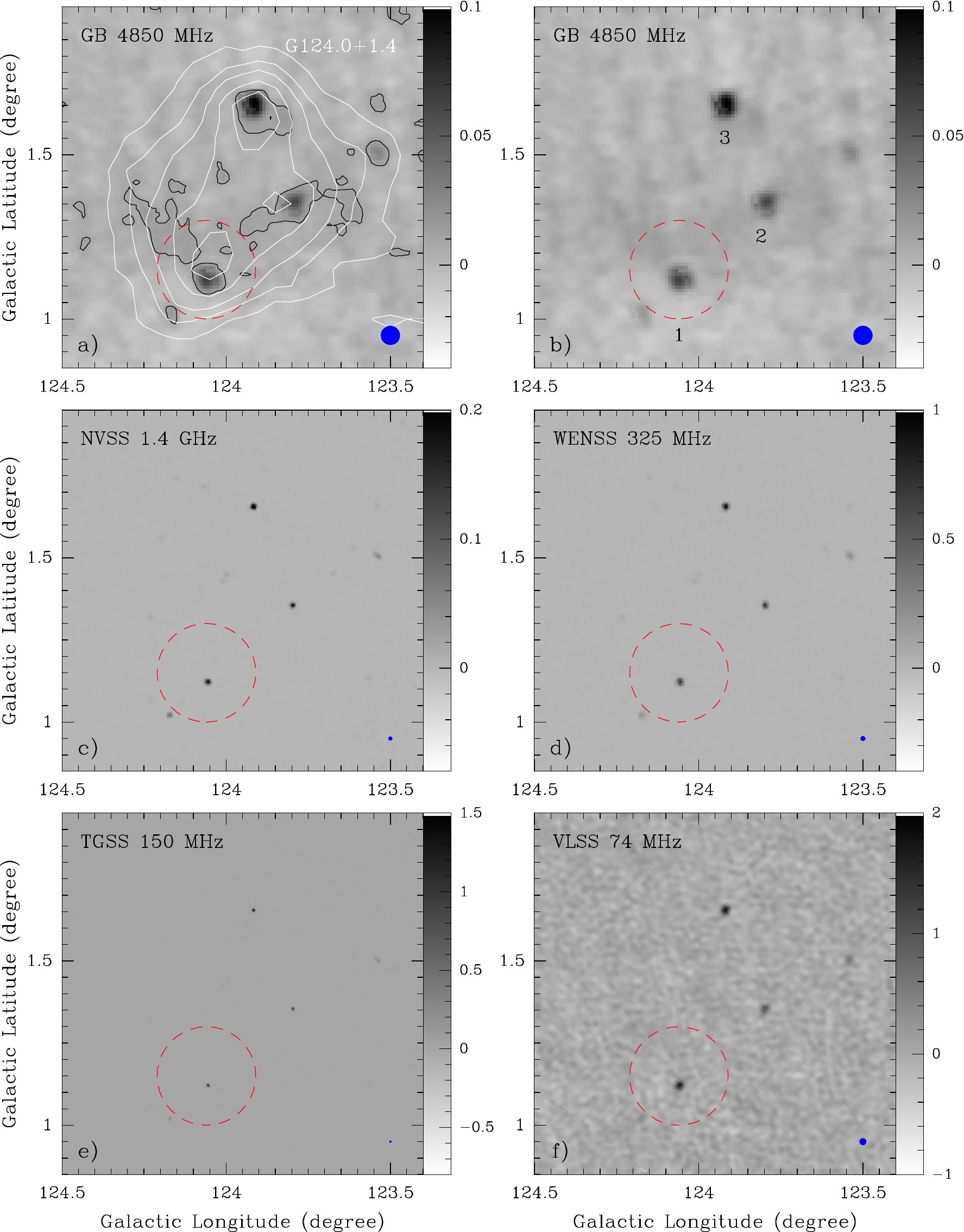}
\end{center}
\caption{\footnotesize The multi-wavelength radio continuum images of G120.0+1.4. The unit of the scale bar is Jy\,beam$^{-1}$ for each panel. The red dashed 
circle shows the bubble center referred from the CO observations, while the blue filled circle at the bottom right corner shows the beam size. (a) The comparison 
between the Green Bank (GB) 4850\,MHz (6\,cm) image (Condon et al. 1994) and Urumqi 6\,cm image (Sun et al. 2007). The GB 4850\,MHz contours represent 
9 mJy emission (about 3\,$\sigma$), while the Urumqi 6\,cm contours (white) represent 5, 10, 15, 22, and 29\,mK emission. (b) The GB 4850\,MHz image. (c) The 
NRAO VLA Sky Survey (NVSS) 1.4\,GHz image (Condon et al. 1998). (d) The Westerbork Northern Sky Survey (WENSS) 325\,MHz image (Rengelink et al. 1997). 
(e) The TIFR GMRT (Giant Metrewave Radio Telescope) Sky Survey (TGSS) 150\,MHz image (Intema et al. 2017). (f) The VLA Low-Frequency Sky Survey (VLSS) 
74\,MHz image (Cohen et al. 2007; Lane et al. 2014).
\label{radio}}
\end{figure*}

\clearpage
\begin{table}[ht!]
\begin{center}
{Tabel~1. The Flux Densities of Radio Sources in G124.0+1.4$^a$}
\footnotesize
\begin{tabular}{l c c c c c c c c} 
\hline\hline 
Object & $l$ & $b$ & 4850\,MHz & 1.4\,GHz & 325\,MHz & 150\,MHz & 74\,MHz & Spectra\\
Name & ($^\circ$) & ($^\circ$)  & (mJy) & (mJy) &  (mJy) &  (mJy) &  (mJy) & Index$^b$ \\ [0.2ex] 
\hline 
Source~1     &   124.05  &   1.12   & 65.8\,$\pm$\,7.8       & 243.5\,$\pm$\,29.2  &  735.8\,$\pm$\,80.9   &  1414.0\,$\pm$\,155.5   &  2052.7\,$\pm$\,246.3   & --0.68\,$\pm$\,0.06 \\ 
Source~2     &   123.79  &   1.35   & 86.5\,$\pm$\,9.9       & 214.3\,$\pm$\,25.7  &  684.4\,$\pm$\,75.3   &  1239.7\,$\pm$\,136.4   &  1319.8\,$\pm$\,158.4   & --0.51\,$\pm$\,0.12 \\ 
Source~3     &   123.92  &   1.65   & 111.3\,$\pm$\,12.6  & 314.2\,$\pm$\,37.7  &  971.9\,$\pm$\,106.9  &   1736.1\,$\pm$\,190.9   &  1926.9\,$\pm$\,231.2  & --0.52\,$\pm$\,0.11 \\ 

\hline
\end{tabular} 
$^a$The flux densities were derived from Gaussian fitting toward the radio sources.\\
$^b$The fitted spectra index $\alpha$ ($S$ $\propto$ $\nu$$^\alpha$) with the flux densities measured between 4850--74\,MHz.\\
\end{center}
\end{table}

\clearpage

\appendix

\section{The Massive Stars in the Shell Region}

\begin{figure}[!ht]
\centering
\includegraphics[scale=0.5]{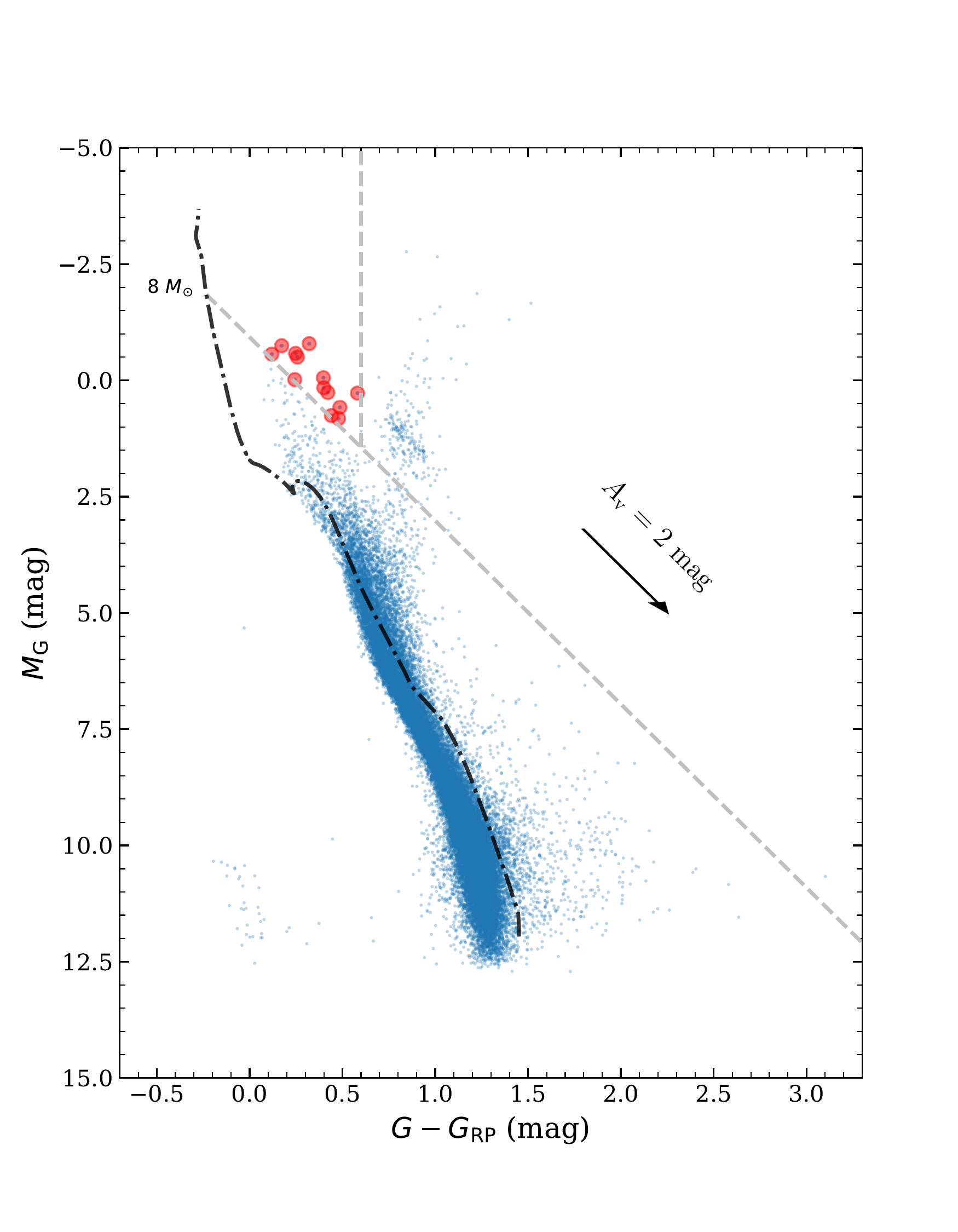}
\caption{Color-magnitude diagram of $Gaia$ EDR3 sources in the shell region. The dash-dotted line represents the $\rm 10~Myr$ isochrone from the PARSEC model (\citealt{Bressan12, Marigo17}). The arrow line is the reddening vector of $A_{\rm v} = \rm{ 2~mag}$, and the gray dashed line parallel to it is the iso-mass line of 
$8~M_{\odot}$. The gray dashed line perpendicular to x-axis is set to distinguish main-sequence stars from giant branch. The blue dots represent the $Gaia$ sources 
and the red-color-filled circles are the identified massive stars.}
\label{cmd}
\end{figure}

\begin{figure}[!ht]
\includegraphics[width=22cm,angle=0]{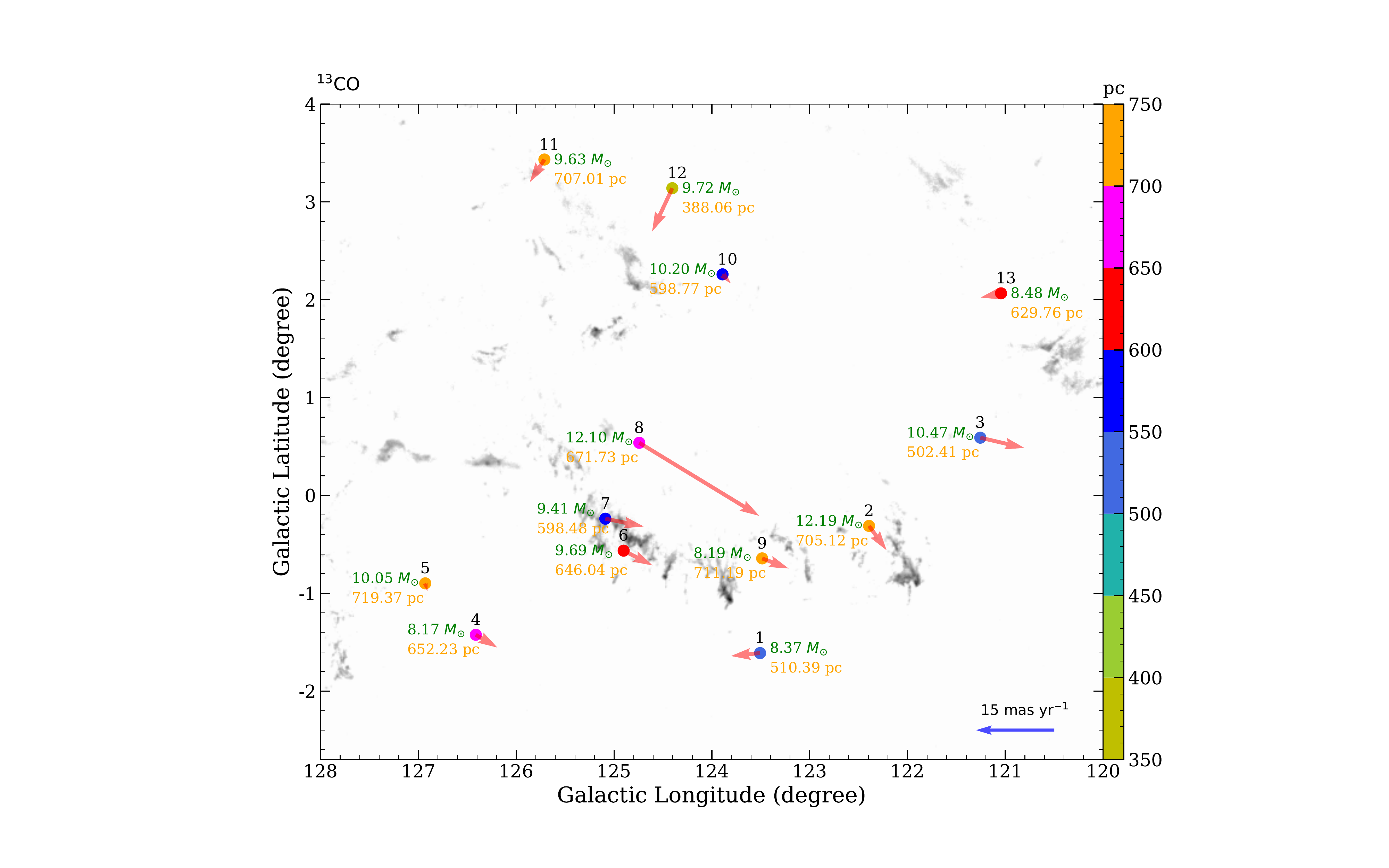}
\caption{The massive stars identified in the shell region by using the $Gaia$ data (see also Table~2), plotted on the $^{13}$CO intensity image. 
For each star, its distance and mass are labelled, while the proper motion is marked by the arrow.}
\label{stars}
\end{figure}

In this Appendix, we try to search for potential massive stars in the shell region by comparing $Gaia$ EDR3 data with the PARSEC model 
(\citealt{Bressan12, Marigo17}). We first apply for the quality cuts to $Gaia$ sources to eliminate the data with high uncertainties. Referring 
to the summary of $Gaia$ EDR3 contents \citep{GaiaCollaboration21}, we set the quality criteria as below:
\begin{flalign}
\begin{split}
&~\rm ruwe < 1.4, \\
&~\varpi_{\rm err} / \varpi < 0.2, \\
&~\rm visibility\_periods\_used > 6, \\
&~\rm {3~mag} < \emph{G} < \rm{21~mag}, \\
&~\rm err\_\emph{G} < 0.01~mag, \\
&~\rm err\_\emph{G}_{\rm RP} < 0.1~mag,
\end{split}& 
\end{flalign}
where ruwe and $\varpi$ represent renormalised unit weight error and parallax, respectively. We then select the sources with distances in the range between 
350 and $\rm 750~pc$, which are considered to be associated with the shell ($d$\,$\sim$\,553~pc). Figure~\ref{cmd} shows the $Gaia$ sources and the $\rm 10~Myr$ 
isochrone in the $G - G_{\rm RP}$ versus $G$ color-magnitude diagram (CMD). The location of each source in the CMD is determined both by its intrinsic color and 
the extinction. Combining the extinction law from \citet{Wang19}, we derive a sample of sources with mass larger than $8~M_{\odot}$. We further remove the giant 
branch from the sample by simply using the criterion of $G - G_{\rm RP} > 0.5$ and consider that the rest are all massive stars. The identified massive stars are 
marked with red circles in Figure~\ref{cmd}. The masses of these stars are obtained from the intersection between the isochrone and the reddening vector, which 
are listed in Table~2. The positions and parallaxes of massive stars are shown in Figure~\ref{stars}.

\clearpage
\begin{table}[ht!]
\begin{center}
{Tabel~2. Massive stars identified in the shell region by using the $Gaia$ data$^a$}
\footnotesize
\begin{tabular}{l c c c c c c c} 
\hline\hline 
Star    & $Gaia$ EDR3       & $l$            & $b$            & Parallax & Mass & PM$_{l}$ & PM$_{b}$ \\
Number  &  Name               & ($^\circ$) & ($^\circ$)  & (mas)      & ($M_{\odot}$) &  (mas/yr) &  (mas/yr)  \\ [0.2ex] 
\hline 
Star~1   &  426670370548306432 &   123.507  &   --1.612   & 1.9593\,$\pm$\,0.0221       & 8.37  &  5.581   &  --0.559  \\ 
Star~2   &  427743322095240576 &   122.393  &   --0.312   & 1.4182\,$\pm$\,0.0168       & 12.19  &  --3.357   &  --4.658  \\
Star~3   &  430852294303966080 &   121.256  &   --0.590   & 1.9904\,$\pm$\,0.0220       & 10.47  &  --8.464   &  --1.969  \\
Star~4   &  510516825178315264 &   126.413  &   --1.425   & 1.5332\,$\pm$\,0.0146       & 8.17  &  --4.133   &  --2.448  \\
Star~5   & 510732295100464128 &   126.930  & --0.899        & 1.3901\,$\pm$\,0.0146      & 10.05  & --0.432    & --1.520   \\
Star~6   & 522682955701497472  &  124.902   & --0.565        & 1.5479\,$\pm$\,0.0148      & 9.69   & --5.526    & --2.818   \\
Star~7  & 523068643761377280  &  125.089   & --0.238        & 1.6709\,$\pm$\,0.0140      & 9.41  &  --7.335   & --1.489   \\
Star~8   & 523305073121434880  &  124.742   & 0.538        & 1.4887\,$\pm$\,0.0107      & 12.10  & --23.008    & --14.009   \\
Star~9  & 523550779611689984  &  123.486   & --0.644        & 1.4061\,$\pm$\,0.0141      & 8.19  & --5.066    & --1.932   \\
Star~10   & 524312053972337024  &  123.891   & 2.260        & 1.6701\,$\pm$\,0.0181      & 10.20  & --1.587    & --1.729   \\
Star~11   & 525362809132485248  & 125.712    & 3.434        & 1.4144\,$\pm$\,0.0201      & 9.63  & 2.782    & --4.319   \\
Star~12   & 525945447214106752  &  124.404   & 3.141        & 2.5769\,$\pm$\,0.0226      & 9.72  & 3.875    & --8.298   \\
Star~13   & 527151538450395776  &  121.044   & 2.066        & 1.5879\,$\pm$\,0.0152      & 8.48  & 3.921    & --0.817   \\
\hline
\end{tabular} 
\end{center}
$^a$The table lists the number (see also Figure~\ref{stars}), $Gaia$ EDR3 name, Galactic position ($l$ and $b$), parallax, mass, and proper motion 
(PM; $l$ and $b$) for each identified star.
\end{table}


\begin{thebibliography}{} 

\bibitem[Abreu-Vicente et al.(2016)]{2016A&A...590A.131A} Abreu-Vicente, J., Ragan, S., Kainulainen, J., et al.\ 2016, \aap, 590, 
A131. doi:10.1051/0004-6361/201527674

\bibitem[Andr{\'e} et al.(2014)]{2014prpl.conf...27A} Andr{\'e}, P., Di Francesco, J., Ward-Thompson, D., et al.\ 2014, Protostars and 
Planets VI, 27. doi:10.2458/azu\_uapress\_9780816531240-ch002

\bibitem[Arthur(2007)]{2007ASSP....1..183A} Arthur, S.~J.\ 2007, in Diffuse Matter from Star Forming Regions to Active Galaxies, 
ed. T. W. Hartquist, J. M. Pittard, \& S. A. E. G. Falle (Berlin: Springer), 183

\bibitem[Ballesteros-Paredes et al.(2020)]{2020SSRv..216...76B} Ballesteros-Paredes, J., Andr{\'e}, P., Hennebelle, P., et al.\ 2020, \ssr, 216, 76. doi:10.1007/s11214-020-00698-3

\bibitem[Ballesteros-Paredes et al.(2011)]{2011MNRAS.411...65B} Ballesteros-Paredes, J., Hartmann, L.~W., V{\'a}zquez-Semadeni, E., et al.\ 2011, \mnras, 411, 65. doi:10.1111/j.1365-2966.2010.17657.x

\bibitem[Ballesteros-Paredes et al.(1999)]{1999ApJ...515..286B} Ballesteros-Paredes, J., V{\'a}zquez-Semadeni, E., \& Scalo, J.\ 1999, \apj, 515, 286. doi:10.1086/307007

\bibitem[Billot et al.(2010)]{2010ApJ...712..797B} Billot, N., Noriega-Crespo, A., Carey, S., et al.\ 2010, \apj, 712, 797. doi:10.1088/0004-637X/712/2/797

\bibitem[Blondin et al.(1996)]{1996ApJ...472..257B} Blondin, J.~M., Lundqvist, P., \& Chevalier, R.~A.\ 1996, \apj, 472, 257. doi:10.1086/178060

\bibitem[Bressan et al.(2012)]{Bressan12} Bressan, A., Marigo, P., Girardi, L., et al.\ 2012, \mnras, 427, 127. doi:10.1111/j.1365-2966.2012.21948.x

\bibitem[Burkert \& Hartmann(2004)]{2004ApJ...616..288B} Burkert, A. \& Hartmann, L.\ 2004, \apj, 616, 288. doi:10.1086/424895

\bibitem[Cazzolato \& Pineault(2003)]{2003AJ....125.2050C} Cazzolato, F. \& Pineault, S.\ 2003, \aj, 125, 2050. doi:10.1086/368242

\bibitem[Chen et al.(2013)]{2013ApJ...769L..16C} Chen, Y., Zhou, P., \& Chu, Y.-H.\ 2013, \apjl, 769, L16. doi:10.1088/2041-8205/769/1/L16

\bibitem[Chevalier \& Blondin(1995)]{1995ApJ...444..312C} Chevalier, R. \& Blondin, J.~M.\ 1995, \apj, 444, 312. doi:10.1086/175606

\bibitem[Chevalier et al.(1992)]{1992ApJ...392..118C} Chevalier, R.~A., Blondin, J.~M., \& Emmering, R.~T.\ 1992, \apj, 392, 118. doi:10.1086/171411

\bibitem[Churchwell(2002)]{2002ARA&A..40...27C} Churchwell, E.\ 2002, \araa, 40, 27. doi:10.1146/annurev.astro.40.060401.093845

\bibitem[Churchwell et al.(2006)]{2006ApJ...649..759C} Churchwell, E., Povich, M.~S., Allen, D., et al.\ 2006, \apj, 649, 759. doi:10.1086/507015

\bibitem[Churchwell et al.(2007)]{2007ApJ...670..428C} Churchwell, E., Watson, D.~F., Povich, M.~S., et al.\ 2007, \apj, 670, 428. doi:10.1086/521646

\bibitem[Cichowolski et al.(2015)]{2015MNRAS.450.3458C} Cichowolski, S., Suad, L.~A., Pineault, S., et al.\ 2015, \mnras, 450, 3458. doi:10.1093/mnras/stv826

\bibitem[Cohen et al.(2007)]{2007AJ....134.1245C} Cohen, A.~S., Lane, W.~M., Cotton, W.~D., et al.\ 2007, \aj, 134, 1245. doi:10.1086/520719

\bibitem[Colombo et al.(2021)]{2021A&A...655L...2C} Colombo, D., K{\"o}nig, C., Urquhart, J.~S., et al.\ 2021, \aap, 655, L2. doi:10.1051/0004-6361/202142182

\bibitem[Condon et al.(1994)]{1994AJ....107.1829C} Condon, J.~J., Broderick, J.~J., Seielstad, G.~A., et al.\ 1994, \aj, 107, 1829. doi:10.1086/116992

\bibitem[Condon et al.(1998)]{1998AJ....115.1693C} Condon, J.~J., Cotton, W.~D., Greisen, E.~W., et al.\ 1998, \aj, 115, 1693. doi:10.1086/300337

\bibitem[Cox et al.(2016)]{2016A&A...590A.110C} Cox, N.~L.~J., Arzoumanian, D., Andr{\'e}, P., et al.\ 2016, \aap, 590, A110. doi:10.1051/0004-6361/201527068

\bibitem[Dale(2015)]{2015NewAR..68....1D} Dale, J.~E.\ 2015, \nar, 68, 1. doi:10.1016/j.newar.2015.06.001

\bibitem[Dale et al.(2015)]{2015MNRAS.450.1199D} Dale, J.~E., Haworth, T.~J., \& Bressert, E.\ 2015, \mnras, 450, 1199. doi:10.1093/mnras/stv396

\bibitem[Dame et al.(2001)]{2001ApJ...547..792D} Dame, T.~M., Hartmann, D., \& Thaddeus, P.\ 2001, \apj, 547, 792. doi:10.1086/318388

\bibitem[Deharveng et al.(2010)]{2010A&A...523A...6D} Deharveng, L., Schuller, F., Anderson, L.~D., et al.\ 2010, \aap, 523, A6. doi:10.1051/0004-6361/201014422

\bibitem[Deharveng et al.(2012)]{2012A&A...546A..74D} Deharveng, L., Zavagno, A., Anderson, L.~D., et al.\ 2012, \aap, 546, A74. doi:10.1051/0004-6361/201219131

\bibitem[de Jager et al.(1988)]{1988A&AS...72..259D} de Jager, C., Nieuwenhuijzen, H., \& van der Hucht, K.~A.\ 1988, \aaps, 72, 259

\bibitem[Dobbs et al.(2014)]{2014prpl.conf....3D} Dobbs, C.~L., Krumholz, M.~R., Ballesteros-Paredes, J., et al.\ 2014, Protostars and Planets VI, 3. 
doi:10.2458/azu\_uapress\_9780816531240-ch001

\bibitem[Du et al.(2017)]{2017ApJS..229...24D} Du, X., Xu, Y., Yang, J., et al.\ 2017, \apjs, 229, 24. doi:10.3847/1538-4365/aa5d9d

\bibitem[Duarte-Cabral \& Dobbs(2017)]{2017MNRAS.470.4261D} Duarte-Cabral, A. \& Dobbs, C.~L.\ 2017, \mnras, 470, 4261. doi:10.1093/mnras/stx1524

\bibitem[Dubner \& Giacani(2015)]{2015A&ARv..23....3D} Dubner, G. \& Giacani, E.\ 2015, \aapr, 23, 3. doi:10.1007/s00159-015-0083-5

\bibitem[Elia et al.(2021)]{2021MNRAS.504.2742E} Elia, D., Merello, M., Molinari, S., et al.\ 2021, \mnras, 504, 2742. doi:10.1093/mnras/stab1038

\bibitem[Elmegreen(1998)]{1998ASPC..148..150E} Elmegreen, B. G. 1998, in ASP Conf. Ser. 148, Origins, ed. C. E. Woodward, 
J. M. Shull, \& H. A. Thronson, Jr. (San Francisco, CA: ASP), 150

\bibitem[Evans(1999)]{1999ARA&A..37..311E} Evans, N.~J.\ 1999, \araa, 37, 311. doi:10.1146/annurev.astro.37.1.311

\bibitem[Evans et al.(2009)]{2009ApJS..181..321E} Evans, N.~J., Dunham, M.~M., J{\o}rgensen, J.~K., et al.\ 2009, \apjs, 181, 321. doi:10.1088/0067-0049/181/2/321

\bibitem[Frerking et al.(1982)]{1982ApJ...262..590F} Frerking, M.~A., Langer, W.~D., \& Wilson, R.~W.\ 1982, \apj, 262, 590. doi:10.1086/160451

\bibitem[Freyer et al.(2003)]{2003ApJ...594..888F} Freyer, T., Hensler, G., \& Yorke, H.~W.\ 2003, \apj, 594, 888. doi:10.1086/376937

\bibitem[Gaia Collaboration et al.(2018)]{2018A&A...616A...1G} Gaia Collaboration, Brown, A.~G.~A., Vallenari, A., et al.\ 2018, \aap, 616, A1. doi:10.1051/0004-6361/201833051

\bibitem[Gaia Collaboration et al.(2021)]{GaiaCollaboration21} Gaia Collaboration, Brown, A.~G.~A., Vallenari, A., et al.\ 2021, \aap, 649, A1. doi:10.1051/0004-6361/202039657

\bibitem[Garmany \& Stencel(1992)]{1992A&AS...94..211G} Garmany, C.~D. \& Stencel, R.~E.\ 1992, \aaps, 94, 211

\bibitem[G{\'o}mez \& V{\'a}zquez-Semadeni(2014)]{2014ApJ...791..124G} G{\'o}mez, G.~C. \& V{\'a}zquez-Semadeni, E.\ 2014, \apj, 791, 124. doi:10.1088/0004-637X/791/2/124

\bibitem[Goodman et al.(2014)]{2014ApJ...797...53G} Goodman, A.~A., Alves, J., Beaumont, C.~N., et al.\ 2014, \apj, 797, 53. doi:10.1088/0004-637X/797/1/53

\bibitem[Guo et al.(2021)]{2021ApJ...921...23G} Guo, W., Chen, X., Feng, J., et al.\ 2021, \apj, 921, 23. doi:10.3847/1538-4357/ac15fe

\bibitem[Hacar et al.(2022)]{2022arXiv220309562H} Hacar, A., Clark, S., Heitsch, F., et al.\ 2022, arXiv:2203.09562

\bibitem[Hacar et al.(2016)]{2016A&A...587A..97H} Hacar, A., Kainulainen, J., Tafalla, M., et al.\ 2016, \aap, 587, A97. doi:10.1051/0004-6361/201526015

\bibitem[Hartmann \& Burkert(2007)]{2007ApJ...654..988H} Hartmann, L. \& Burkert, A.\ 2007, \apj, 654, 988. doi:10.1086/509321

\bibitem[Hennebelle(2013)]{2013A&A...556A.153H} Hennebelle, P.\ 2013, \aap, 556, A153. doi:10.1051/0004-6361/201321292

\bibitem[Heyer \& Dame(2015)]{2015ARA&A..53..583H} Heyer, M. \& Dame, T.~M.\ 2015, \araa, 53, 583. doi:10.1146/annurev-astro-082214-122324

\bibitem[Heyer et al.(2009)]{2009ApJ...699.1092H} Heyer, M., Krawczyk, C., Duval, J., et al.\ 2009, \apj, 699, 1092. doi:10.1088/0004-637X/699/2/1092

\bibitem[Intema et al.(2017)]{2017A&A...598A..78I} Intema, H.~T., Jagannathan, P., Mooley, K.~P., et al.\ 2017, \aap, 598, A78. doi:10.1051/0004-6361/201628536

\bibitem[Inutsuka et al.(2015)]{2015A&A...580A..49I} Inutsuka, S.-. ichiro ., Inoue, T., Iwasaki, K., et al.\ 2015, \aap, 580, A49. doi:10.1051/0004-6361/201425584

\bibitem[Inutsuka \& Miyama(1997)]{1997ApJ...480..681I} Inutsuka, S.-. ichiro . \& Miyama, S.~M.\ 1997, \apj, 480, 681. doi:10.1086/303982

\bibitem[Jackson et al.(2010)]{2010ApJ...719L.185J} Jackson, J.~M., Finn, S.~C., Chambers, E.~T., et al.\ 2010, \apjl, 719, L185. doi:10.1088/2041-8205/719/2/L185

\bibitem[Jackson et al.(2021)]{2021AAS...23713702J} Jackson, J., Whitaker, J., Chambers, E., et al.\ 2021, \aas

\bibitem[Kauffmann et al.(2013)]{2013ApJ...779..185K} Kauffmann, J., Pillai, T., \& Goldsmith, P.~F.\ 2013, \apj, 779, 185. doi:10.1088/0004-637X/779/2/185

\bibitem[Krumholz et al.(2014)]{2014prpl.conf..243K} Krumholz, M.~R., Bate, M.~R., Arce, H.~G., et al.\ 2014, Protostars and Planets VI, 243. doi:10.2458/azu\_uapress\_9780816531240-ch011

\bibitem[Krumholz et al.(2019)]{2019ARA&A..57..227K} Krumholz, M.~R., McKee, C.~F., \& Bland-Hawthorn, J.\ 2019, \araa, 57, 227. doi:10.1146/annurev-astro-091918-104430

\bibitem[Kudritzki \& Puls(2000)]{2000ARA&A..38..613K} Kudritzki, R.-P. \& Puls, J.\ 2000, \araa, 38, 613. doi:10.1146/annurev.astro.38.1.613

\bibitem[Landecker et al.(1992)]{1992A&A...258..495L} Landecker, T.~L., Anderson, M.~D., Routledge, D., et al.\ 1992, \aap, 258, 495

\bibitem[Lane et al.(2014)]{2014MNRAS.440..327L} Lane, W.~M., Cotton, W.~D., van Velzen, S., et al.\ 2014, \mnras, 440, 327. doi:10.1093/mnras/stu256

\bibitem[Larson(1981)]{1981MNRAS.194..809L} Larson, R.~B.\ 1981, \mnras, 194, 809. doi:10.1093/mnras/194.4.809

\bibitem[Li et al.(2016)]{2016A&A...591A...5L} Li, G.-X., Urquhart, J.~S., Leurini, S., et al.\ 2016, \aap, 591, A5. doi:10.1051/0004-6361/201527468

\bibitem[Li et al.(2013)]{2013A&A...559A..34L} Li, G.-X., Wyrowski, F., Menten, K., et al.\ 2013, \aap, 559, A34. doi:10.1051/0004-6361/201322411

\bibitem[Marigo et al.(2017)]{Marigo17} Marigo, P., Girardi, L., Bressan, A., et al.\ 2017, \apj, 835, 77. doi:10.3847/1538-4357/835/1/77

\bibitem[Mattern et al.(2018)]{2018A&A...619A.166M} Mattern, M., Kauffmann, J., Csengeri, T., et al.\ 2018, \aap, 619, A166. doi:10.1051/0004-6361/201833406

\bibitem[McClure-Griffiths et al.(2002)]{2002ApJ...578..176M} McClure-Griffiths, N.~M., Dickey, J.~M., Gaensler, B.~M., et al.\ 2002, \apj, 578, 176. doi:10.1086/342470

\bibitem[McKee \& Ostriker(2007)]{2007ARA&A..45..565M} McKee, C.~F. \& Ostriker, E.~C.\ 2007, \araa, 45, 565. doi:10.1146/annurev.astro.45.051806.110602

\bibitem[Miceli et al.(2013)]{2013MNRAS.430.2864M} Miceli, M., Orlando, S., Reale, F., et al.\ 2013, \mnras, 430, 2864. doi:10.1093/mnras/stt093

\bibitem[Molinari et al.(2010)]{2010A&A...518L.100M} Molinari, S., Swinyard, B., Bally, J., et al.\ 2010, \aap, 518, L100. doi:10.1051/0004-6361/201014659

\bibitem[Motte et al.(2018)]{2018ARA&A..56...41M} Motte, F., Bontemps, S., \& Louvet, F.\ 2018, \araa, 56, 41. doi:10.1146/annurev-astro-091916-055235

\bibitem[Myers(2009)]{2009ApJ...700.1609M} Myers, P.~C.\ 2009, \apj, 700, 1609. doi:10.1088/0004-637X/700/2/1609

\bibitem[Ostriker(1964)]{1964ApJ...140.1056O} Ostriker, J.\ 1964, \apj, 140, 1056. doi:10.1086/148005

\bibitem[Qin et al.(2008)]{2008A&A...484..361Q} Qin, S.-L., Wang, J.-J., Zhao, G., et al.\ 2008, \aap, 484, 361. doi:10.1051/0004-6361:20078483

\bibitem[Ragan et al.(2014)]{2014A&A...568A..73R} Ragan, S.~E., Henning, T., Tackenberg, J., et al.\ 2014, \aap, 568, A73. doi:10.1051/0004-6361/201423401

\bibitem[Reid et al.(2019)]{2019ApJ...885..131R} Reid, M.~J., Menten, K.~M., Brunthaler, A., et al.\ 2019, \apj, 885, 131. doi:10.3847/1538-4357/ab4a11

\bibitem[Rengelink et al.(1997)]{1997A&AS..124..259R} Rengelink, R.~B., Tang, Y., de Bruyn, A.~G., et al.\ 1997, \aaps, 124, 259. doi:10.1051/aas:1997358

\bibitem[Schneider et al.(2016)]{2016A&A...591A..40S} Schneider, N., Bontemps, S., Motte, F., et al.\ 2016, \aap, 591, A40. doi:10.1051/0004-6361/201628328

\bibitem[Schneider \& Elmegreen(1979)]{1979ApJS...41...87S} Schneider, S. \& Elmegreen, B.~G.\ 1979, \apjs, 41, 87. doi:10.1086/190609

\bibitem[Shan et al.(2012)]{Shan2012} Shan, W.~L., Yang, J., Shi, S.~C., et al. 2012, IEEE Transactions on Terahertz Science and Technology, 2, 593

\bibitem[Shimajiri et al.(2019)]{2019A&A...623A..16S} Shimajiri, Y., Andr{\'e}, P., Palmeirim, P., et al.\ 2019, \aap, 623, A16. doi:10.1051/0004-6361/201834399

\bibitem[Simpson et al.(2012)]{2012MNRAS.424.2442S} Simpson, R.~J., Povich, M.~S., Kendrew, S., et al.\ 2012, \mnras, 424, 2442. doi:10.1111/j.1365-2966.2012.20770.x

\bibitem[Smith et al.(2014)]{2014MNRAS.441.1628S} Smith, R.~J., Glover, S.~C.~O., Clark, P.~C., et al.\ 2014, \mnras, 441, 1628. doi:10.1093/mnras/stu616

\bibitem[Smith et al.(2020)]{2020MNRAS.492.1594S} Smith, R.~J., Tre{\ss}, R.~G., Sormani, M.~C., et al.\ 2020, \mnras, 492, 1594. doi:10.1093/mnras/stz3328

\bibitem[Soam et al.(2017)]{2017MNRAS.465..559S} Soam, A., Maheswar, G., Lee, C.~W., et al.\ 2017, \mnras, 465, 559. doi:10.1093/mnras/stw2649

\bibitem[Solomon et al.(1987)]{1987ApJ...319..730S} Solomon, P.~M., Rivolo, A.~R., Barrett, J., et al.\ 1987, \apj, 319, 730. doi:10.1086/165493

\bibitem[Sousbie(2011)]{2011MNRAS.414..350S} Sousbie, T.\ 2011, \mnras, 414, 350. doi:10.1111/j.1365-2966.2011.18394.x

\bibitem[Su et al.(2019)]{2019ApJS..240....9S} Su, Y., Yang, J., Zhang, S., et al.\ 2019, \apjs, 240, 9. doi:10.3847/1538-4365/aaf1c8

\bibitem[Suad et al.(2014)]{2014A&A...564A.116S} Suad, L.~A., Caiafa, C.~F., Arnal, E.~M., et al.\ 2014, \aap, 564, A116. doi:10.1051/0004-6361/201323147

\bibitem[Sun et al.(2007)]{2007A&A...463..993S} Sun, X.~H., Han, J.~L., Reich, W., et al.\ 2007, \aap, 463, 993. doi:10.1051/0004-6361:20066001

\bibitem[Sun et al.(2011)]{2011A&A...536A..83S} Sun, X.~H., Reich, P., Reich, W., et al.\ 2011, \aap, 536, A83. doi:10.1051/0004-6361/201117693

\bibitem[Sun et al.(2021)]{2021ApJS..256...32S} Sun, Y., Yang, J., Yan, Q.-Z., et al.\ 2021, \apjs, 256, 32. doi:10.3847/1538-4365/ac11fe

\bibitem[Tutone et al.(2020)]{2020A&A...642A..67T} Tutone, A., Orlando, S., Miceli, M., et al.\ 2020, \aap, 642, A67. doi:10.1051/0004-6361/202038336

\bibitem[Wang \& Chevalier(2002)]{2002ApJ...574..155W} Wang, C.-Y. \& Chevalier, R.~A.\ 2002, \apj, 574, 155. doi:10.1086/340795

\bibitem[Wang et al.(2016)]{2016ApJS..226....9W} Wang, K., Testi, L., Burkert, A., et al.\ 2016, \apjs, 226, 9. doi:10.3847/0067-0049/226/1/9

\bibitem[Wang et al.(2015)]{2015MNRAS.450.4043W} Wang, K., Testi, L., Ginsburg, A., et al.\ 2015, \mnras, 450, 4043. doi:10.1093/mnras/stv735

\bibitem[Wang \& Chen(2019)]{Wang19} Wang, S. \& Chen, X.\ 2019, \apj, 877, 116. doi:10.3847/1538-4357/ab1c61

\bibitem[Wareing et al.(2017)]{2017MNRAS.470.2283W} Wareing, C.~J., Pittard, J.~M., \& Falle, S.~A.~E.~G.\ 2017, \mnras, 470, 2283. doi:10.1093/mnras/stx1417

\bibitem[Watson et al.(2010)]{2010ApJ...716.1478W} Watson, C., Hanspal, U., \& Mengistu, A.\ 2010, \apj, 716, 1478. doi:10.1088/0004-637X/716/2/1478

\bibitem[Weaver et al.(1977)]{1977ApJ...218..377W} Weaver, R., McCray, R., Castor, J., et al.\ 1977, \apj, 218, 377. doi:10.1086/155692

\bibitem[Whalen, \& Norman(2008)]{2008ApJ...672..287W} Whalen, D.~J., \& Norman, M.~L.\ 2008, \apj, 672, 287

\bibitem[Wilson \& Rood(1994)]{1994ARA&A..32..191W} Wilson, T.~L. \& Rood, R.\ 1994, \araa, 32, 191. doi:10.1146/annurev.aa.32.090194.001203

\bibitem[Wright(2020)]{2020NewAR..9001549W} Wright, N.~J.\ 2020, \nar, 90, 101549. doi:10.1016/j.newar.2020.101549

\bibitem[Xu et al.(2021)]{2021A&A...645L...8X} Xu, Y., Hou, L.~G., Bian, S.~B., et al.\ 2021, \aap, 645, L8. doi:10.1051/0004-6361/202040103

\bibitem[Yan et al.(2021)]{2021A&A...645A.129Y} Yan, Q.-Z., Yang, J., Sun, Y., et al.\ 2021, \aap, 645, A129. doi:10.1051/0004-6361/202039768

\bibitem[Yuan et al.(2021)]{2021ApJS..257...51Y} Yuan, L., Yang, J., Du, F., et al.\ 2021, \apjs, 257, 51. doi:10.3847/1538-4365/ac242a

\bibitem[Zhang et al.(2016)]{2016A&A...585A.117Z} Zhang, C.-P., Li, G.-X., Wyrowski, F., et al.\ 2016, \aap, 585, A117. doi:10.1051/0004-6361/201526296

\bibitem[Zhang et al.(2019)]{2019A&A...622A..52Z} Zhang, M., Kainulainen, J., Mattern, M., et al.\ 2019, \aap, 622, A52. doi:10.1051/0004-6361/201732400

\bibitem[Zhou et al.(2014)]{2014ApJ...791..109Z} Zhou, X., Yang, J., Fang, M., et al.\ 2014, \apj, 791, 109. doi:10.1088/0004-637X/791/2/109

\bibitem[Zucker et al.(2015)]{2015ApJ...815...23Z} Zucker, C., Battersby, C., \& Goodman, A.\ 2015, \apj, 815, 23. doi:10.1088/0004-637X/815/1/23

\bibitem[Zucker et al.(2018)]{2018ApJ...864..153Z} Zucker, C., Battersby, C., \& Goodman, A.\ 2018, \apj, 864, 153. doi:10.3847/1538-4357/aacc66

\bibitem[Zucker et al.(2022)]{2022Natur.601..334Z} Zucker, C., Goodman, A.~A., Alves, J., et al.\ 2022, \nat, 601, 334. doi:10.1038/s41586-021-04286-5

\bibitem[Zucker et al.(2019)]{2019ApJ...887..186Z} Zucker, C., Smith, R., \& Goodman, A.\ 2019, \apj, 887, 186. doi:10.3847/1538-4357/ab517d

\end{thebibliography}
\end{document}